\documentclass[aps,pra,
superscriptaddress,
reprint,twocolumn,preprintnumbers,
amsmath,amssymb,
nofootinbib]{revtex4-1}

\usepackage[titletoc,toc,title]{appendix}
\usepackage{braket}
\usepackage{comment}
\usepackage{amsmath}
\usepackage{color}
\usepackage{graphicx}
\usepackage[font=small]{subcaption}
\usepackage[font=small]{caption}

\makeatletter
\def\endfmffile{%
	\fmfcmd{\p@rcent\space the end.^^J%
			end.^^J%
			endinput;}%
	\if@fmfio
		\immediate\closeout\@outfmf
	\fi
	\ifnum\pdfshellescape=\@ne
		\immediate\write18{mpost \thefmffile}%
	\fi}
\makeatother

\usepackage{graphicx}
\usepackage{dcolumn}
\usepackage{bm}
\usepackage{hyperref}%

\newcommand{\beq}{\begin{equation}}
\newcommand{\eeq}{\end{equation}}

\begin{document}


\title{Glassy quantum dynamics in translation invariant fracton models}

\author{Abhinav Prem}
\email{abhinav.prem@colorado.edu}
\affiliation{
Department of Physics and Center for Theory of Quantum Matter,
University of Colorado, Boulder, Colorado 80309, USA
}
\author{Jeongwan Haah}
\affiliation{
Station Q Quantum Architectures and Computation Group,
Microsoft Research, Redmond, Washington, USA
}
\author{Rahul Nandkishore}
\affiliation{
Department of Physics and Center for Theory of Quantum Matter,
University of Colorado, Boulder, Colorado 80309, USA
}

\begin{abstract}
We investigate relaxation in the recently discovered ``fracton" models 
and discover that these models naturally host glassy quantum dynamics in the absence of quenched disorder.
We begin with a discussion of ``type~I" fracton models, 
in the taxonomy of Vijay, Haah, and Fu. 
We demonstrate that in these systems, the mobility of charges 
is suppressed {\it exponentially} in the inverse temperature. 
We further demonstrate that when a zero temperature type~I fracton model is placed 
in contact with a finite temperature heat bath, 
the approach to equilibrium is a {\it logarithmic} function of time 
over an exponentially wide window of time scales. 
Generalizing to the more complex ``type~II" fracton models, 
we find that the charges exhibit {\it subdiffusion} 
upto a relaxation time that diverges at low temperatures as a {\it super-exponential} function of inverse temperature. 
This behaviour is reminiscent of ``nearly localized" disordered systems, 
but occurs with a translation invariant three-dimensional Hamiltonian. 
We also conjecture that fracton models with conserved charge 
may support a phase which is a {\it thermal} metal but a {\it charge} insulator. 
\end{abstract}

\maketitle


\section{Introduction and Motivation}
\label{intro}

Recent years have witnessed an explosion of interest 
in many body localization (MBL)~\cite{anderson1958, Mirlin, BAA, Prosen, PalHuse, Imbrie, ARCMP},
whereby isolated quantum systems with quenched disorder can exhibit ergodicity breaking 
and fail to thermalize even at infinite times. 
Interest in the phenomenon is in part intrinsic 
(e.g. many body localized systems can support entirely new types of quantum order~\cite{
LPQO, VoskAltman2014, Pekkeretal2014, MBL116}),
in part practical 
(e.g. MBL systems can serve as ideal quantum memories~\cite{ARCMP}),
and in part due to the tantalizing connections to other fields. 
For example, MBL can be viewed as a type of ``ideal quantum glass" which does not reach equilibrium at infinite times---can insights from MBL inform the study of classical structural glass? (This particular question has been the focus of intensive work~\cite{MaksimovKagan, Grover2014, Schiulazbubbles, yaoglass, Papicstoudenmire, deroeck1, deroeck2, Garrahanglass} without any definitive conclusions). Recent experimental progress~\cite{MBL92, kondov, bloch1, bloch2, bordia, NMR3, Lukin} has further intensified interest in the field. However, theoretical insights have been hard won, due to the fundamental challenges of describing a non-ergodic, non-equilibrium, strongly correlated, highly disordered phase. Indeed most theoretical progress has required either a new idea (such as the notion of emergent integrability~\cite{Serbynlbits, HNO, Ros, Imbrie, GeraedtsBhattNandkishore}), a new technique (such as dynamical versions of real space renormalization~\cite{MBL93})---or a new class of models which provide access to a new and formerly unforeseen phenomenology (e.g.~\cite{vpp}).

The study of topological phases in three spatial dimensions 
has brought to light a new class of 
models~\cite{Chamon2005,Bravyi2011,Haah2011,Yoshida2013,Vijay2015,Vijay2016} 
which have remarkable properties. 
These exactly solvable three-dimensional (3D) lattice models 
have ground states that exhibit a sub-extensive topological degeneracy 
on the 3D torus and possess point-like excitations, 
dubbed ``fractons"~\cite{Vijay2015} 
that cannot move without creating additional excitations. 
Such systems are related by duality to spin models 
with symmetries along lower-dimensional subsystems~\cite{Vijay2016,Williamson2016} 
and were recently classified into Type~I and Type~II fracton phases. 
In the Type~I phases, fractons are created 
by the application of a membrane operator 
and pairs of fractons form composite topological excitations 
that can move along lower-dimensional subsystems. 
In the Type~II phases, 
fractons are created by the application of a fractal operator 
and all topological excitations are strictly immobile. 
The fracton models that have been introduced to date 
have all involved discrete symmetries, 
although there does not in principle appear to be any obstruction to constructing continuous 
fracton models---indeed a stimulating series of works from Pretko~\cite{Pretko1, Pretko2} 
appears to reproduce much of the fracton phenomenology 
within a continuum field theory with $U(1)$ symmetry. 
Layer constructions of these phases have also recently been advanced \cite{Vijay2017,Ma2017}. 

In this work, we combine emerging ideas from research into (fracton) topological phases and MBL by studying the {\it dynamical} behaviour of fracton models, revealing an intimate and provocative connection between the two fields. We begin by discussing type~I fracton models at finite energy density and demonstrate that in these models, the mobility of charges is suppressed {\it exponentially} in the inverse temperature. When a type~I fracton model prepared in its ground space (i.e. at zero energy density) is placed in contact with a finite temperature heat bath, we show that the equilibration exhibits $\log t$ behaviour over an exponentially wide window of time scales---a classic signature of glassy dynamics (see e.g.~\cite{Amir} and references contained therein). We emphasize that this glassy quantum dynamics occurs in a {\it three-dimensional, translation invariant} Hamiltonian. We then turn to type~II fracton models, and demonstrate that charges exhibit subdiffusion upto a relaxation time that diverges at low temperatures as a {\it super-exponential} function of temperature, reminiscent of ``near MBL" disordered systems~\cite{GN, MirlinMuller}. Finally, we conjecture that fracton models with conserved $U(1)$ charge could realize exotic three-dimensional phases that are thermal metals but charge insulators. 

Our work has striking implications for MBL, for the study of fracton phases, and for possible technological applications of both. For MBL, our work illuminates new connections to glasses, introducing a new class of models that exhibit glassy quantum dynamics with translation invariant Hamiltonians. It may also inform investigations of localization and glassy dynamics in higher dimensions~\cite{2dcontinuum, avalanches, MirlinMuller}. For the field of fracton phases, our work reveals that not only do these models have an unusual ground state structure, they also support rich quantum dynamics, thus opening a new line of investigation for three-dimensional topological phases. Practically speaking, our work also uncovers a new route to information storage, as well as identifying a potential class of three-dimensional phases that are thermal metals but charge insulators. These last could have applications e.g. in high density electronics, where the problem of heat dissipation currently constrains possibilities for miniaturization.

\section{Fracton Models}
\label{fracton}

Fracton topological phases are a new class of three-dimensional phases of matter that display features that go beyond those familiar from gauge theory. These phases can be obtained as the quantum duals of three-dimensional systems with symmetries along lower dimensional sub-systems, specifically along planes and fractals. A unified framework, based on a generalized lattice gauge theory, for fracton topological order was recently proposed in~\cite{Vijay2016}. Given the novelty of these phases, in this section we provide a self-contained exposition of fracton systems, focusing primarily on specific examples to elucidate the features most relevant to the dynamics. 

Fracton phases arise in exactly solvable lattice models in three spatial dimensions and exhibit a sub-extensive topological ground state degeneracy on the 3D torus. The distinguishing feature of these systems is the presence of point-like fractional excitations---fractons---that are fundamentally immobile i.e., they cannot move without creating additional topological excitations. In contrast with anyons in two-dimensional topologically ordered systems, where anyons are created at the ends of a Wilson line and are thus allowed to move by application of a local line-like operator, there exists no local line-like operator that creates a pair of fractons. Instead, fractons are created at the ends of membrane or fractal operators, leaving a single fracton immobile. A classification scheme for fracton topological order was recently proposed in~\cite{Vijay2016}, where these systems were divided into type~I and type~II phases.

Type I fracton phases, such as the X-Cube model discussed below, host fracton excitations at the ends of membrane operators. While single fractons are immobile, bound-states of fractons form composite topological excitations that are free to move along lower-dimensional subsystems such as a line or a plane. There may also exist additional quasi-particles that are confined to move only along lower-dimensional subsystems. In type~II phases, such as Haah's code~\cite{Haah2011}, fractons are created by the application of fractal operators and \emph{all} topological excitations are strictly immobile. The latter feature leads to a fundamental difference between the dynamics of type~I and II fracton phases, which we now consider separately in the following sections.

\section{Type I Fracton Models}
\label{type1}

\begin{figure}[b]
\includegraphics[width=8cm]{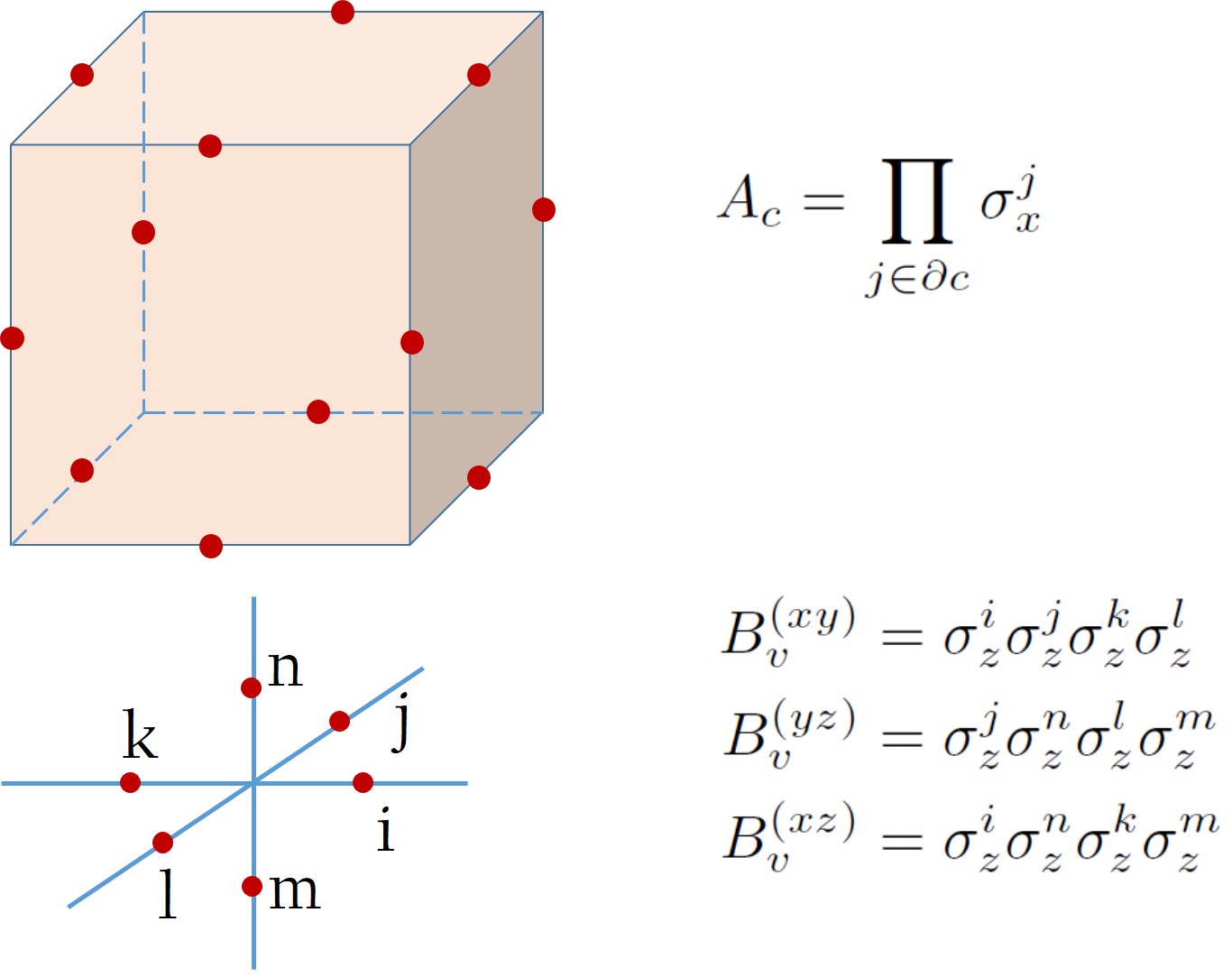}
\centering
\caption{
\raggedright 
The X-Cube model is represented by spins $\sigma$ placed on the links of a cubic lattice and is given by the sum of a twelve-spin $\sigma_x$ operator at each cube $c$ and planar four-spin $\sigma_z$ operators at each vertex $v$.}
\label{XCube}
\end{figure}

The physics of type I fracton topological order is best illustrated through the example of the X-Cube model~\cite{Vijay2016} that displays the essential features of these phases. The X-Cube model is an exactly solvable lattice model defined on a cubic lattice with Ising spins living on each link. The Hamiltonian is 
\begin{align}
H_{XC} = - \sum_c A_c - \sum_{v,k} B_c^{(k)} ,
\label{HXC}
\end{align}
where the first term is the sum over all cubes of a twelve-spin $\sigma_x$ interaction 
and the second term is the sum over all vertices of planar four-spin $\sigma_z$ interactions 
as depicted in Fig.~\ref{XCube}. In contrast with two-dimensional topologically ordered
states, such as the Toric Code, that have a finite and constant topological ground state 
degeneracy on the two-torus, the ground state of the pure X-Cube model on the three-torus 
of linear dimension $L$ has a sub-extensive topological ground state degeneracy $D$, where 
$\log_2 D = 6L - 3$.

\begin{figure}[b]
\centering
\begin{subfigure}[b]{0.23\textwidth}
\includegraphics[width=\textwidth]{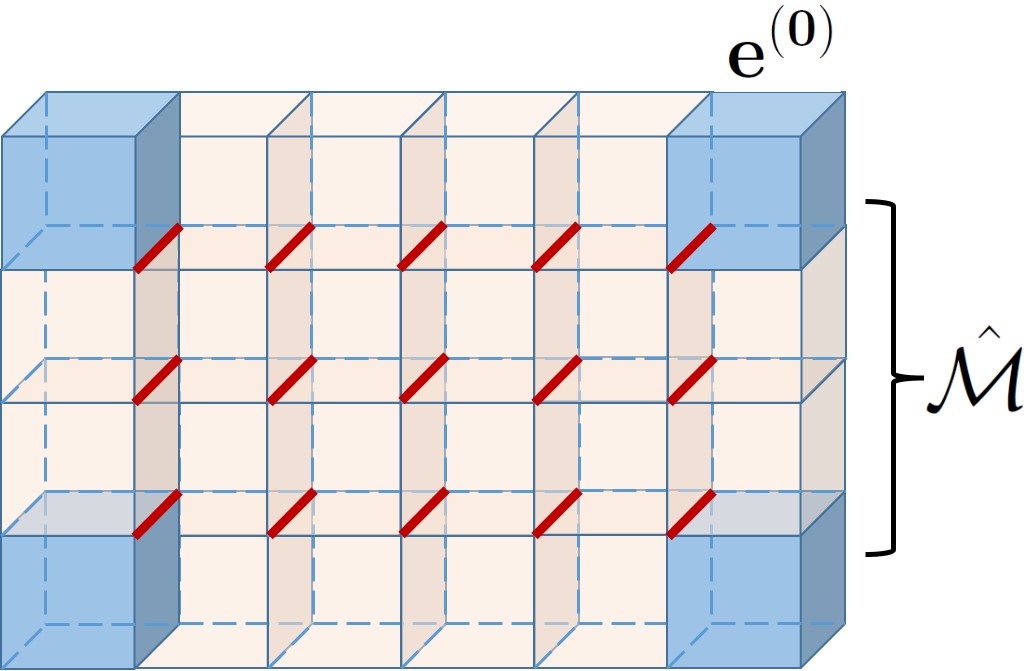}
\caption{}
\label{membrane-1}
\end{subfigure}\quad\quad
\begin{subfigure}[b]{0.21\textwidth}
\includegraphics[width=\textwidth]{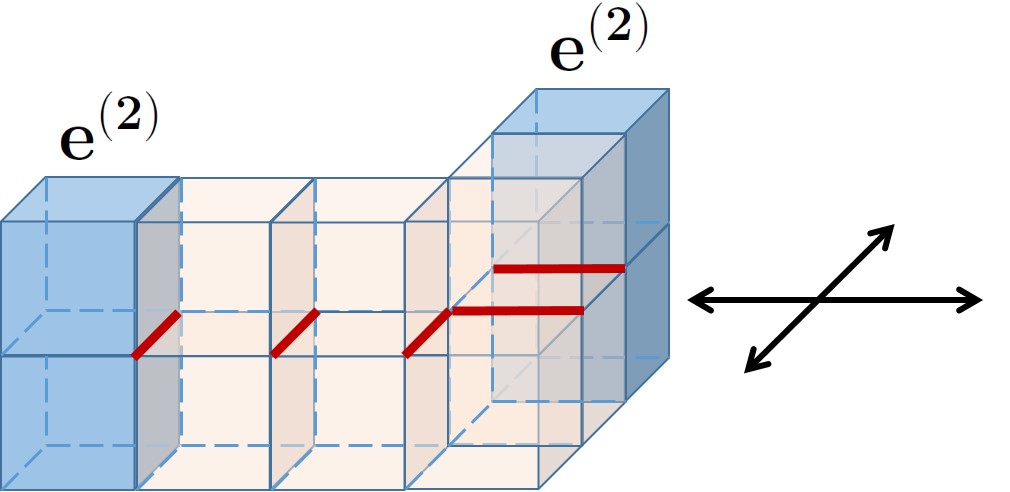}
\caption{}
\label{membrane-2}
\end{subfigure}
\caption{\raggedright 
Topological excitations of the X-Cube model are depicted in (i) and (ii). 
Fractons $e^{(0)}$ are created at corners by acting on the ground state 
by a membrane operator $\mathcal{M}$ that is the product of $\sigma_z$ operators along red links. 
Wilson line operators
create a composite topological excitation $e^{(2)}$. }
\end{figure}

A fracton is created by flipping the eigenvalue of the cubic interaction term. 
However, there exists no local operator 
that can create a single pair of fractons. 
Indeed, applying a $\sigma_z$ operator to a link 
flips the eigenvalues of the four cubes sharing that link. 
Acting on the ground state by $\sigma_z$ along a membrane operator $\hat{\mathcal{M}}$ 
creates four fractons at the corners of the membrane, as shown in Fig.~\ref{membrane-1}.
A single fracton, denoted $e^{(0)}$ 
(where the superscript denotes that it is a dimension-0 excitation), 
is thus fundamentally immobile, 
as moving it would create additional fractons. 
This is the fundamental ``superselection" rule \cite{Kim2015}
that will lead to glassy dynamics. 
Pairs of fractons are however free to move by repeated application of local membrane operators. 
A straight Wilson line of $\sigma_z$ operators 
creates a pair of fractons at each end---each pair is a composite excitation 
that can move in two dimensions, and which we refer to as a dimension-2 (dim-2) excitation $e^{(2)}$, 
as shown in Fig.~\ref{membrane-2}. 
In the X-Cube model, there exist additional dimension-1 excitations ($m^{(1)}$)
created at the ends of a Wilson line of $\sigma_x$ operators, that are mobile 
along one-dimensional sub-manifolds. 
Henceforth we will refer to the fully mobile four-fracton composites, created 
by single $\sigma_z$ operators, as the topologically neutral sector, and the lower 
dimensional excitations (fractons and $e^{(2)}$'s) as the topologically charged sector. 

In the following sections, we will focus specifically on the X-Cube model in the presence 
of transverse fields,
\beq
H = - J \sum_c A_c - \sum_{v,k} B_c^{(k)} + \Lambda \sum_i \sigma_z + \lambda \sum_i \sigma_x ,
\label{HXCfull}
\eeq
where $i$ goes over all links in the cubic lattice. Since the pure X-Cube model Eq.~\eqref{HXC} is a sum of commuting projectors, the relative coefficient $J$ simply sets the energy scale between the $e$ and $m$ excitations when $\Lambda,\lambda = 0$. In the presence of the transverse fields, we expect that the fracton phase will survive up to some finite $\Lambda/J$ and $\lambda/J$ since this is a gapped phase of matter that is stable to local perturbations~\cite{BravHast2010,BravHast2011}. In the limit of large transverse fields however, the fracton topological order will be destroyed, but the precise nature of the transition between the fracton phase and the trivial paramagnetic phase has yet to be understood~\cite{Vijay2016}. Since we are interested in the dynamics within the fracton phase, allowing only for weak local perturbations, we set $J = 1$, noting that our analysis holds as long $J$ is $O(1)$. In addition, we will first set $\lambda = 0$ and consider only the dynamics of the fractons and their composites. After analysing this sector, we will comment on the consequences of a non-zero $\lambda$, which would allow the $m^{(1)}$ particles to hop as well. The Hamiltonian pertinent for the following discussions is thus
\beq
H = - \sum_c A_c - \sum_{v,k} B_c^{(k)} + \Lambda \sum_i \sigma_z ,
\label{HXCmain}
\eeq
with the perturbation strength $\Lambda \ll 1$. We note that while we are focusing on the 
specific example of the X-Cube model, the results presented here hold broadly 
for all type~I fracton phases.%
\footnote{
The X-Cube model lacks an exact electromagnetic duality, 
present in other type~I models e.g., the Majorana cube or Checkerboard models, 
which have two distinct fracton excitations, $e^{(0)}$ and $m^{(0)}$. 
However, we can always choose to initialize the system 
with only one topological excitation 
or perturb the system only by terms that allow one of the excitations to hop.}

\subsection{Type I Fractons at Finite Energy Density}
\label{finiteenergy}

We begin our discussion of dynamics in fracton models 
by considering the X-Cube model Eq.~\eqref{HXCmain} at 
finite energy density. Since we have switched off the term
that would allow $m$'s to hop, we have three kinds of 
excitations with dynamics in our system: neutral composites, 
which, being fully mobile and created by local terms, are three-dimensional bosons; 
dim-2 excitations $e^{(2)}$, which are two-dimensional bosons~\cite{Vijay2016}; 
and the topologically charged fractons. A word on notations---since both the 
neutral composites and $e^{(2)}$'s are bosonic, we will henceforth refer to the 
former as the composite ($c$) sector and the latter as the bosonic ($b$) sector, 
with fractons ($f$) sometimes also referred to as the (topologically) charged sector. 
 
\begin{figure}[b]
\centering
\includegraphics[width=8cm]{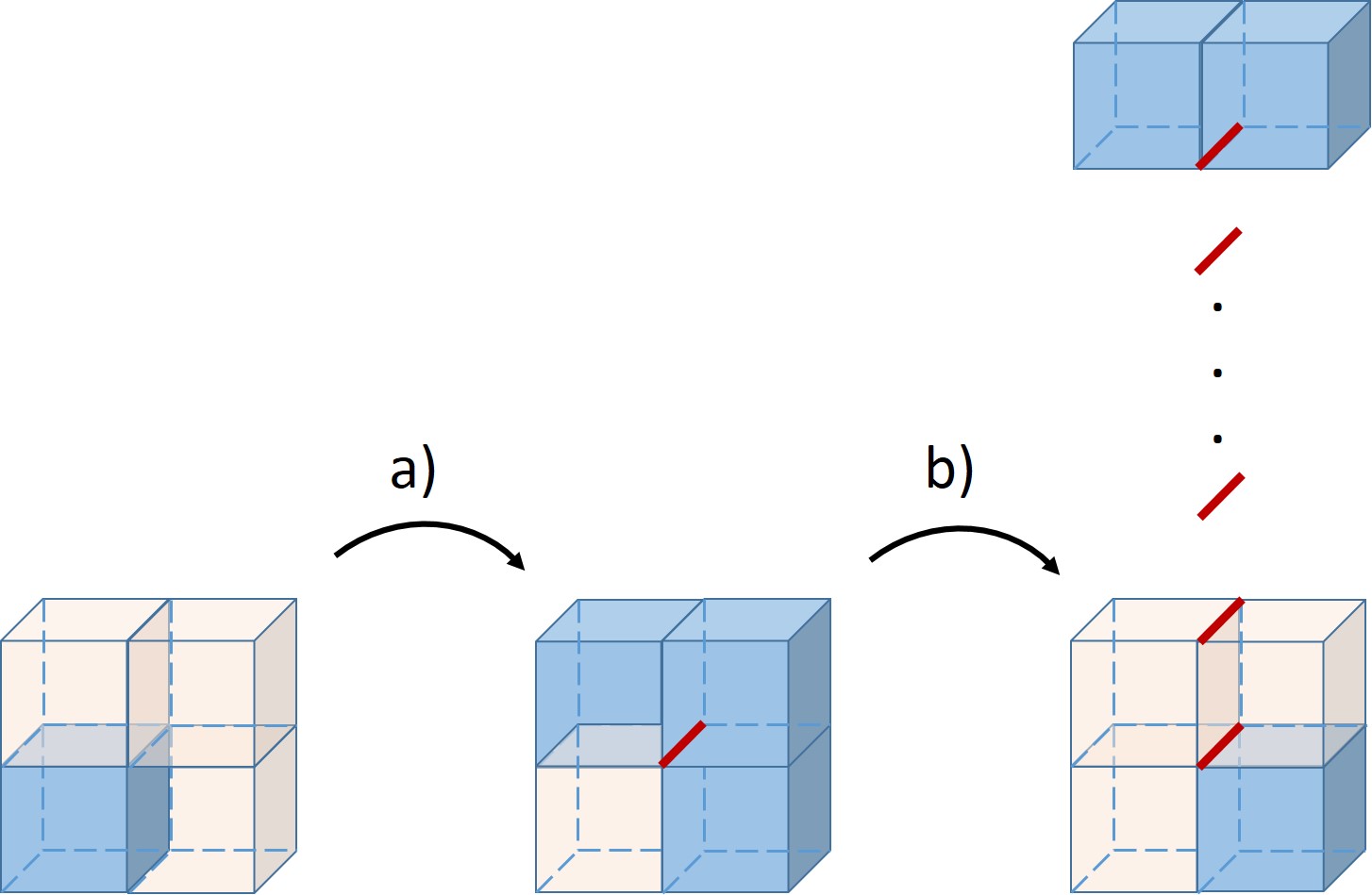}
\caption{\raggedright A single fracton hop. Starting from a single isolated fracton, we can move it over by one site by the action of a $\sigma_z$ operator, shown in red, in step a). However, this creates two additional fractons which together form a dimension-1 excitation that can move along a line without creating any further excitations. As shown in b), this pair can then be moved off to infinity by the action of a Wilson line of $\sigma_z$ operators.}
\label{hop}
\end{figure}

From the preceding general discussion of the model, 
it is clear that each fracton hop 
is accompanied by the creation of two additional fractons, 
and so energetically costs an amount $W = 4$. We will henceforth refer to this charge gap simply as $W$, since the analysis applies also to generalizations of Eq.~\eqref{HXCmain} that are in the same phase, but perhaps with a different charge gap. 
A single fracton hop is depicted in Fig.~\ref{hop}: 
starting with an isolated fracton, 
we can move this over by one site by acting by a single $\sigma_z$ operator 
on the link adjacent to the fracton. 
This creates two additional topological excitations; 
however, this pair is a dim-2 $e^{(2)}$ excitation and can be moved off to infinity 
at no additional energy cost. 
Thus, each hop takes the system off energy shell. To fully understand the relaxation 
in the fracton sector, we must thus take into account the fully mobile neutral 
composites and the dim-2 excitations. These sectors act as a thermalizing bath 
for the fractons as rearrangements within these sectors allow the system 
to come back on energy shell after each hop.

Before studying the non-equilibrium dynamics, we consider the system---comprised of 
fractons, $e^{(2)}$'s, and neutral composites---in equilibrium, at some temperature $T \ll W$. The 
local energy scale in the topologically charged sector is $W$, where $W$ is the charge 
gap, i.e., the cost of creating two fractons. Further, the fracton sector is coupled to 
a dense, high temperature bath of neutral composites which hop at a rate $\Lambda \ll W$. Here 
each species is gapped, with a gap of $W/2$, $W$, and $2 W$ for creating a single 
fracton, $e^{(2)}$, and composite respectively. Additionally, since the composites 
and $e^{(2)}$'s are neutral, with the fractons carrying only a $\mathbb{Z}_2$ charge, 
the density and temperature of these excitations cannot be controlled independently; 
rather, the equilibrium temperature uniquely determines the density of each species, 
\beq
n_f \sim e^{-\frac{W}{2 T}}, \quad n_b \sim e^{-\frac{W}{T}}, \quad n_c \sim e^{-\frac{2W}{T}},
\eeq
where $n_f$ is the density of fractons, $n_b$ is the density of dim-2 bosonic 
excitations, and $n_c$ is the density of the composites.

Due to the form of the perturbation, $\sigma_z$, the only processes that lead to 
an exchange of energy between the three sectors are those where the total $\mathbb{Z}_2$ 
charge along each row and each column of the cubic lattice is conserved. For instance, in 
Fig.~\ref{hop}, the initial and final topological charges along each column and each row 
are preserved (modulo 2). From all possible on-shell processes by which the three sectors 
can exchange energy, we will consider only two body processes 
that are up to second order in $\Lambda$ (shown in Figs.~\ref{bosoncomposite} and \ref{bosonfracton}) 
since all others will be further suppressed either 
by density factors or by the perturbation strength. As an illustration, Fig.~\ref{bosoncomposite} 
depicts the processes where a composite breaks into two $e^{(2)}$'s 
and where two $e^{(2)}$'s combine into a 
composite. 

Within the thermalizing bath, the composites and $e^{(2)}$'s can either hop 
at a rate $\Lambda$ or scatter off each other while remaining locally on-shell, 
at a rate $\Lambda^2/W$. Since we are interested in 
the regime where $\Lambda \ll W$, the heat bath has a narrow local bandwidth 
$\sim \Lambda$, determined primarily by the hopping. As discussed earlier, a single 
fracton hop takes the system off energy shell by an amount $W$, and hence a single rearrangement 
within the bath cannot place the system on shell. The traditional analysis of localized systems 
coupled to narrow bandwidth baths~\cite{GN, Parameswaran2016, General} 
makes use of many body rearrangements in the charge sector 
to obtain a relaxation rate that is power law slow in the bandwidth of the bath. 
However, in that setting the charge sector admits local re-arrangements 
that are ``uphill" or ``downhill," 
which may be combined into a many body re-arragement 
that is off shell by much less than $W$.
The present situation differs in that {\it every} local re-arrangement in the charge sector 
is ``uphill" in energy, 
so the dominant relaxation mechanism from Ref.~\cite{GN, Parameswaran2016, General} does not apply.
Since the maximum energy the bath can provide is $\tilde{\Lambda} = \min(\Lambda, T)$, the bath must be probed 
$n \sim W/\tilde{\Lambda}$ times 
for the fractons to borrow enough energy to perform a single hop. 
This leads to a relaxation rate in the charge sector that is {\it exponentially} slow in $W/\tilde{\Lambda}$,
scaling as 
\begin{align}
\label{expslowrelaxation}
\Gamma \sim n_f e^{-n} \sim n_f e^{-W/\tilde{\Lambda}}.
\end{align}
Alternatively, we could make use of the extensive nature of the many body bandwidth in the bath 
and could use an $n\sim W/\tilde{\Lambda}$ particle rearrangement in the bath to place fracton rearrangements on shell.
However, this yields~\cite{slowheating} an exponentially slow relaxation of the same form as 
Eq.~\eqref{expslowrelaxation}, up to subleading prefactors. 

\begin{figure}[b]
\centering
\includegraphics[width=8cm]{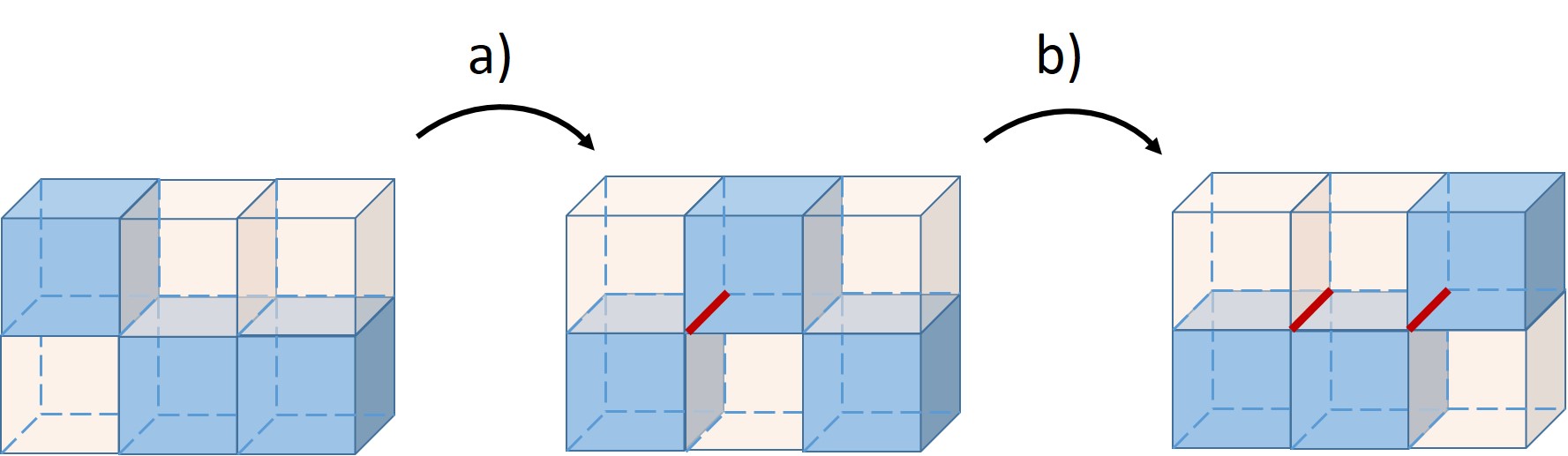}
\caption{\raggedright A fracton hop mediated by an $e^{(2)}$ excitation. a) Starting from a single isolated fracton and an $e^{(2)}$, we first act by a $\sigma_z$ operator (in red) resulting in three fractons. b) Acting with another $\sigma_z$ takes us to a configuration with an $e^{(2)}$ and a fracton that has moved over by two sites. In this manner, fractons can hop while remaining on-shell.}
\label{bosonhop}
\end{figure}

The fractons, however, have an additional channel through which they can hop in an on-shell manner. 
This process, depicted in Fig.~\ref{bosonhop}, requires the presence of an $e^{(2)}$ excitation in 
the vicinity of the fracton. During this process, the fracton hops once, destroying the neighbouring 
$e^{(2)}$, and hops again, thereby returning the system on-shell by creating an $e^{(2)}$ particle. 
Since this process is mediated by the $e^{(2)}$'s, the rate at which it 
proceeds is additionally suppressed by the density of the bosons $n_b$
\beq
\Gamma \sim n_f n_b \sim n_f\,e^{-W/T},
\eeq
Thus, the fractons have two possible hopping channels---either by borrowing 
energy from the bath or by using the bosons as intermediaries. The faster 
rate will dominate and so 
\beq
\label{relax}
\Gamma \sim n_f \max\left(e^{-W/T}, e^{-W/\tilde{\Lambda}}\right) \sim n_f\,e^{-W/T}.
\eeq
This behaviour is in marked contrast with the usual case of activated transport. In typical 
gapped phases, the relaxation displays an Arrhenius law $\Gamma_A \sim n \sim e^{-\Delta/T}$, where 
$n$ is the density of charge carries and $\Delta$ is the charge gap. This exponential slowness of 
the relaxation is governed primarily by the exponential rarity of charge carriers, since the mobility 
of these excitations is generically $\sim O(1)$. 
In type I fracton models, while the relaxation is 
suppressed by the density of fractons $n_f$,
which is exponentially small in the inverse temperature,
it displays an additional suppression due to mobility,
which is also exponentially small in the inverse temperature.
This is conceptually different from Arrhenius 
behaviour. Additionally, we observe that insofar as $T$ is the potentially ergodicity restoring parameter 
here and the relaxation rate is exponentially small therein, 
Eq.~\eqref{relax} corresponds to ``asymptotic many body localization" 
in the taxonomy of Ref.~\cite{deroeck1, deroeck2}. 

We note that hole burning, 
a scenario where, after a certain number of fracton hops, 
there is a depletion in the density of composites 
causing the remaining fractons to essentially be frozen 
until the composite sector thermalizes, does not arise here. 
This is due to the exponentially long time scales, 
$t \sim e^{W/T}$, 
over which the composite bath is probed, 
allowing it to effectively thermalize between probing attempts. 
Hole burning would only occur if the energy ``consumption" $W \Gamma$ is 
larger than the energy ``supply" $T$
\begin{align}
W e^{-W/T} > T ,
\end{align}
which does not hold true for a low temperature bath.

We note that we have thus far neglected the ``back action" of the fractons on the composite sector.
Insofar as $W/T \gg 1$, 
the back action is strong, 
and it thus may be tempting to postulate that 
the composite sector itself could be localized by the coupling to the charge sector, 
in a form of MBL proximity effect~\cite{proximity, MBL118, General}. 
If the bath gets localized, 
then the relaxation rate will be zero.
However, there will inevitably be regions 
where the density of fractons is low enough for the composite sector to be locally ergodic, 
and this argument will thus inevitably run into the rare region obstruction to perturbative constructions 
that have derailed previous attempts to establish translation invariant MBL~\cite{Schiulazbubbles}. 
We therefore do not pursue this particular line of reasoning further, 
noting only that Eq.~\eqref{expslowrelaxation} 
should properly be understood only as an upper bound on the relaxation rate, 
which could be slower because of back action on the bath. 

The situation considered above is one where the three sectors are in equilibrium. We could, 
in principle, prepare the system out of equilibrium with the bath comprised only of the composites 
and not the dim-2 $e^{(2)}$'s. In this case, the hopping would proceed at a rate
\beq
\Gamma \sim n_f\,e^{-W/\tilde{\Lambda}},
\eeq
since the channel where hopping proceeds through an intermediate boson will be unavailable. Hole burning 
will again be avoided here as we have a narrow bandwidth bath at high temperature. However, 
as we show in App.~\ref{bathdetails}, there will be a rapid equilibration between the composite and bosonic 
sectors, and upon equilibration, the relaxation will revert to the rate $n_f\,e^{-W/T}$.

\subsection{Glassy dynamics in the approach to equilibrium}
\label{fractonglass}

We now consider the non-equilibrium dynamics of type~I fracton models 
prepared in their ground state (i.e., at vanishing energy density)
in contact with a low but finite temperature bath comprised of 
neutral composites (the fully mobile bosons) and dim-2 
excitations (the two-dimensional bosons). 
Let us begin our discussion more generally, 
by initializing the fractons at a temperature $T_f^{(0)}$ 
such that the density of fractons $n_f \sim e^{-W/2 T_f^{(0)}}$,
where $W$ is again the local energy scale in the fracton sector. 
The neutral composites are prepared at a temperature $T_c^{(0)}$ and the 
$e^{(2)}$'s are prepared at $T_b^{(0)}$, with corresponding 
densities $n_c \sim e^{-2W/T_c^{(0)}}$ and $n_b \sim e^{-W/T_b^{(0)}}$.
Furthermore, since all three species are gapped here, every species has an 
exponentially small heat capacity, $C_i \sim n_i$, for each species $i = b,c,f$. 
We note that the usual low-temperature 
$T^3$ heat capacity that we expect for three-dimensional bosons holds only 
when the bosons are gapless and thus does not apply to the neutral composites. 
Similarly, for the dim-2 bosons the usual $T^2$ behaviour which 
follows from the Stefan-Boltzmann law does not hold here since that 
applies only to gapless bosons. Hence, here we expect that
\begin{align}
\label{heatcap}
E_c \sim 2 W n_c &\implies C_c \sim \frac{4 W^2}{T_c^2} e^{-\frac{2 W}{T_c}}, \nonumber \\
E_b \sim W n_b &\implies C_b \sim \frac{W^2}{T_b^2} e^{-\frac{W}{T_b}}, \nonumber \\
E_f \sim \frac{W}{2} n_f &\implies C_f \sim \frac{W^2}{4 T_f^2} e^{-\frac{W}{2T_f}}.
\end{align}
The regime of interest is $T_i^{(0)} \ll \Lambda$ where $i=c,b,f$,
and we henceforth work in this regime. We further show in the Appendix~\ref{DiffEq}
that even if we start with $T_c\neq T_b$ the boson and composite sectors rapidly equilibrate,
so we henceforth assume that $T_c=T_b$.
That is, we assume that the bath is itself in equilibrium, 
and examine the equilibration of the fracton sector with the bath. 

\subsubsection{Equilibration between Fractons and bath}

\begin{figure}[t]
\centering
\begin{subfigure}[b]{0.5\textwidth}
\includegraphics[width=\textwidth]{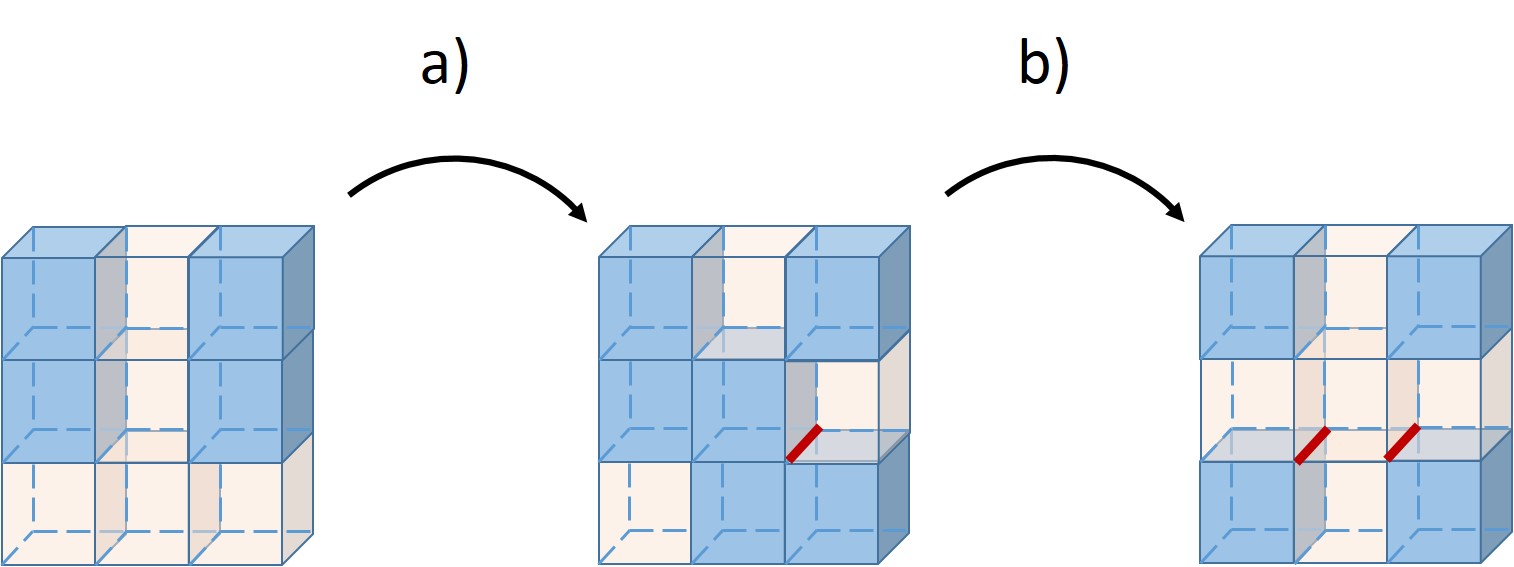}
\caption{}
\label{bosonfracton1}
\end{subfigure}\\
\begin{subfigure}[b]{0.5\textwidth}
\includegraphics[width=\textwidth]{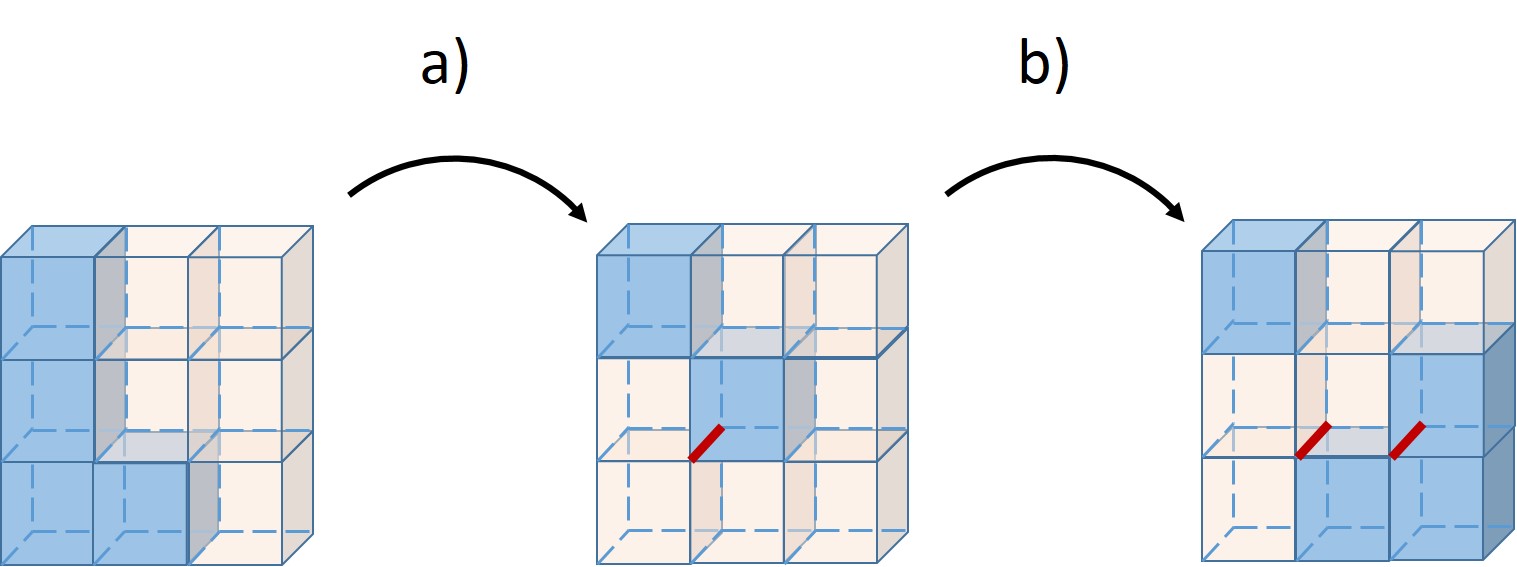}
\caption{}
\label{bosonfracton2}
\end{subfigure}\\
\begin{subfigure}[b]{0.5\textwidth}
\includegraphics[width=\textwidth]{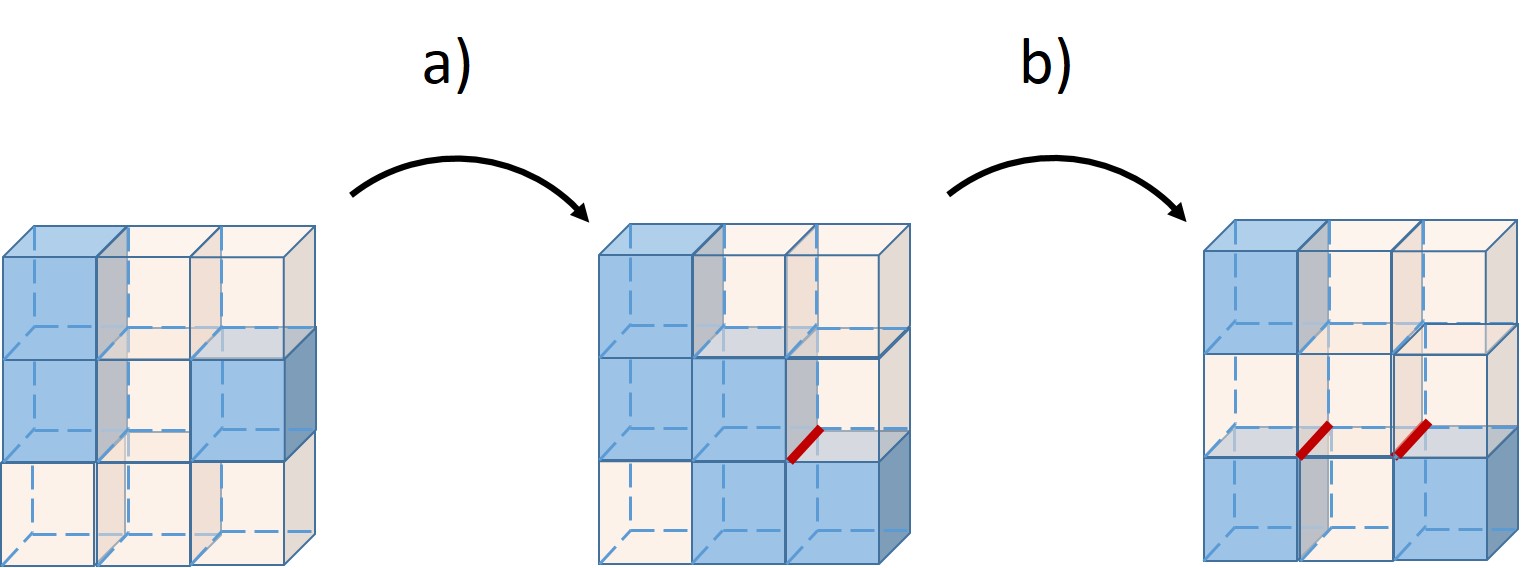}
\caption{}
\label{bosonfracton3}
\end{subfigure}
\caption{\raggedright 
Dominant second order on-shell processes between charged fracton sector and thermal bath. 
(i) Two bosons convert into four fractons. 
(ii) Two bosons convert into a boson and two fractons. 
(iii) A boson and a fracton convert into three fractons.}
\label{bosonfracton}
\end{figure}

We now consider the dynamics of the fractons when placed in contact 
with the thermal sector which contains composites and $e^{(2)}$'s at some initial temperature 
$T_b^{(0)}$. The fractons are prepared at a low temperature, $T_f^{(0)} \ll T_b^{(0)}$, as 
we are interested in the dynamics of fractons prepared in their ground state. Further, 
we will consider only two-body (processes involving more excitations will be 
further suppressed by density factors) 
on-shell processes (upto second order in perturbation 
theory) that lead to an exchange of energy between the bath and the $\mathbb{Z}_2$-charged fracton 
sector. Since we are assuming that the composites and bosons have already equilibrated, 
the dominant processes through which the bath and the $\mathbb{Z}_2$-charged fracton 
sector exchange energy will involve only the bosons and the fractons. This is because the 
density of neutral composites $\sim e^{-2W/T_b}$ while that of the bosons $\sim e^{-W/T_b}$, which dominates 
in the regime of interest $T_b \ll W$. The dominant processes, shown in Fig.~\ref{bosonfracton}, are then
\begin{enumerate}
\item Boson + Boson $\leftrightarrow$ 4 Fractons,
\item Boson + Boson $\leftrightarrow$ Boson + 2 Fractons,
\item Boson + Fracton $\leftrightarrow$ 3 Fractons.
\end{enumerate}

Two fractons cannot convert into a single boson due to the form of the perturbation, 
and all other two body processes will involve at least one composite and are hence 
suppressed compared to the others. Processes where a single composite converts into a 
pair of bosons which then convert into four fractons or a boson and two fractons occur at
third order in perturbation theory and can hence safely be ignored. 
The above three channels all occur at second order and their rates are controlled by the densities of 
the excitations involved. For instance, channel 1 (boson + boson) occurs at a rate $ \Gamma \sim 
\frac{\Lambda^2}{W} n_b^2$. 
Since the bath lends
energy $2 W$ during this process, 
$dE_b/dt \sim -2 W \Gamma = -2 \Lambda^2 n_b$. 
Following this example, we can establish a 
detailed balance between the thermal and the charged sector,
\begin{align}
\frac{d E_b}{d t} =  - \Lambda^2 \left(3 n_b^2 + n_b n_f - n_b n_f^2 - n_f^3 - 2 n_f^4\right) = -\frac{d E_f}{d t}.
\end{align}
Here, the integer coefficients' magnitude should not be taken seriously,
since they are strongly model dependent. However, the conclusions we draw below will not 
depend on these coefficients, lending our results broader applicability across type I fracton phases.

Since the heat capacities are given by Eq.~\eqref{heatcap}, the detailed balance leads to the rate equations
\begin{align}
\label{fractonboson}
\frac{d T_b}{d t} &=  - \frac{\Lambda^2 T_b^2}{W^2} \left(3 n_b + n_f - n_f^2 - \frac{n_f^3}{n_b} - 2 \frac{n_f^4}{n_b}\right), \nonumber \\
\frac{d T_f}{d t} &= \frac{4\Lambda^2 T_f^2}{W^2} \left(3 \frac{n_b^2}{n_f} + n_b - n_b n_f - n_f^2 - 2 n_f^3\right).
\end{align}
We can analytically study the equilibration process in two regimes---when the fractons are prepared in their ground space ($T_f \sim 0$) and when the system nears equilibration ($T_b \sim T_f$).

\begin{figure}[t]
\centering
\includegraphics[width=8cm]{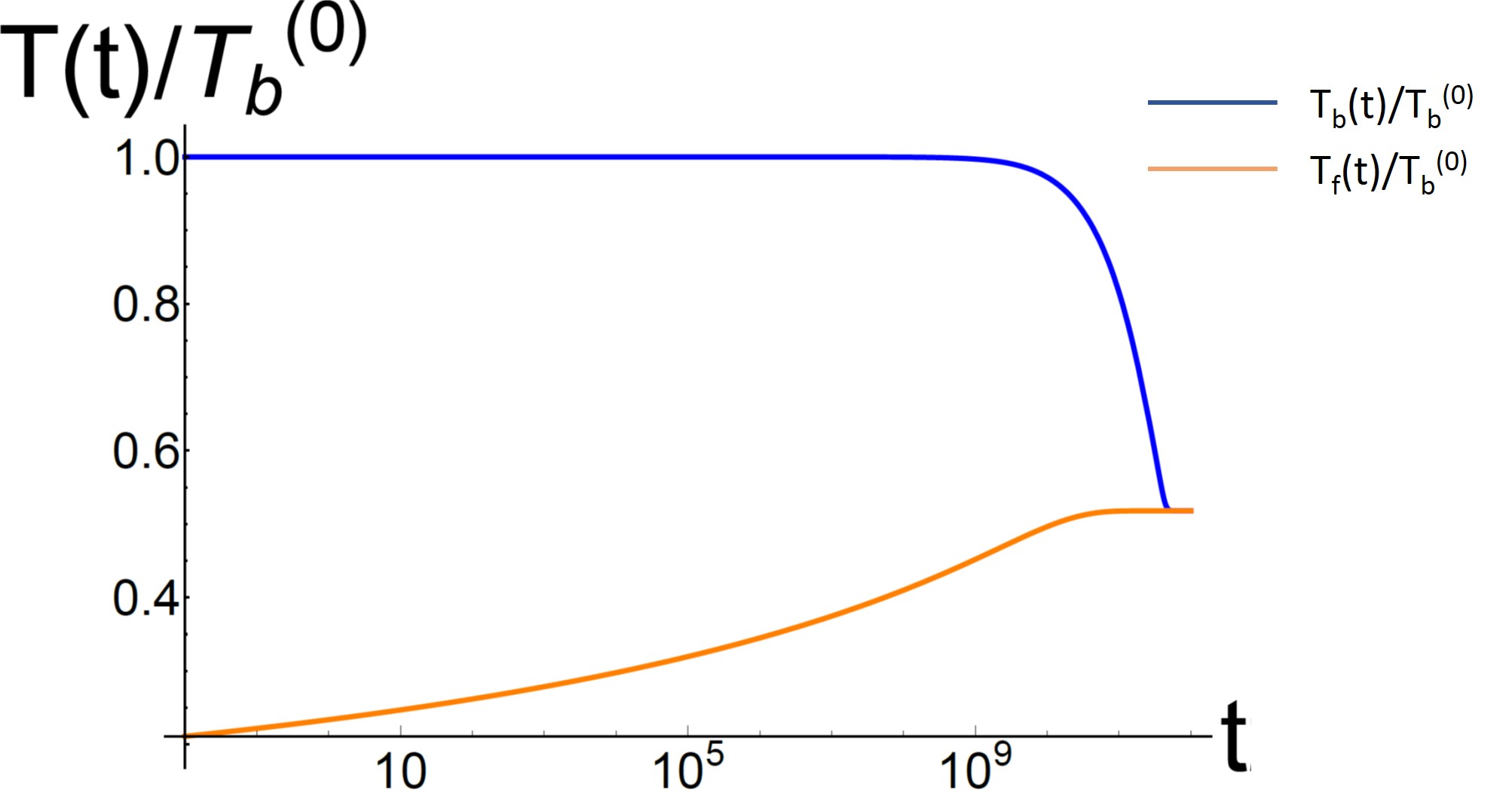}
\caption{\raggedright Time dependence of the charged fracton $T_f(t)$ and thermal $T_b(t)$ sectors found by numerically solving the detailed balance equation, Eq.~\eqref{fractonboson}. We have initialized the system at $W=1$, $\Lambda = 10^{-1}$, $T_b^{(0)} = 5 \times 10^{-2}$, and $T_f^{(0)} = 5 \times 10^{-4}$.}
\label{tempplot}
\end{figure}

\begin{figure}[t]
\centering
\includegraphics[width=8cm]{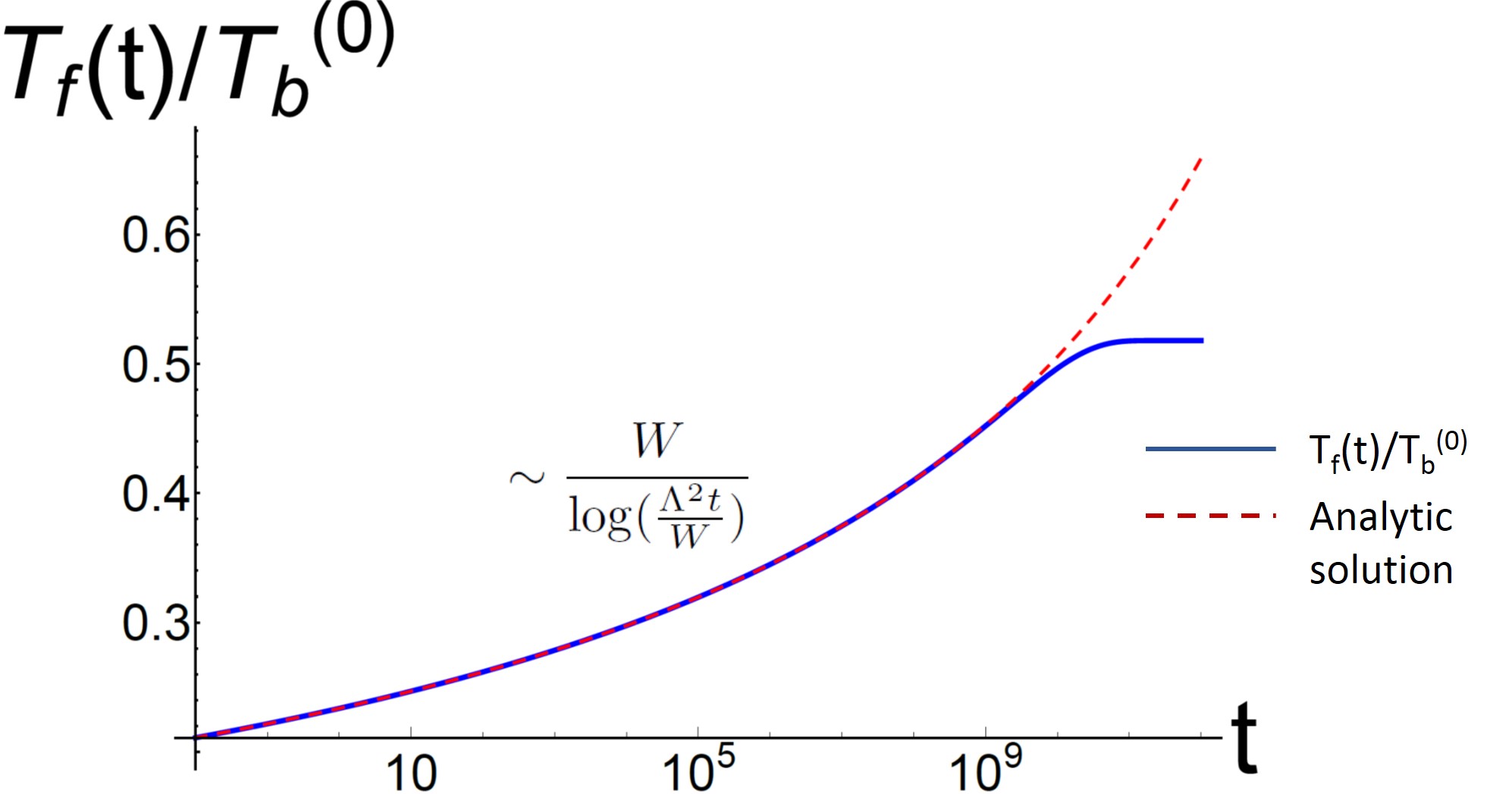}
\caption{\raggedright The time dependence of the fracton sector $T_f(t)$ displays a $\log(t)$ behaviour over an exponentially long time scale. The red dashed line is the analytic approximation, Eq.~\eqref{fractontemp} to the solution of Eq.~\eqref{fractonboson}, with the numerical solution shown in blue. The same parameters are used as those in Fig.~\ref{tempplot}.}
\label{tempfracton}
\end{figure}

\begin{figure}[tbp]
\centering
\includegraphics[width=8cm]{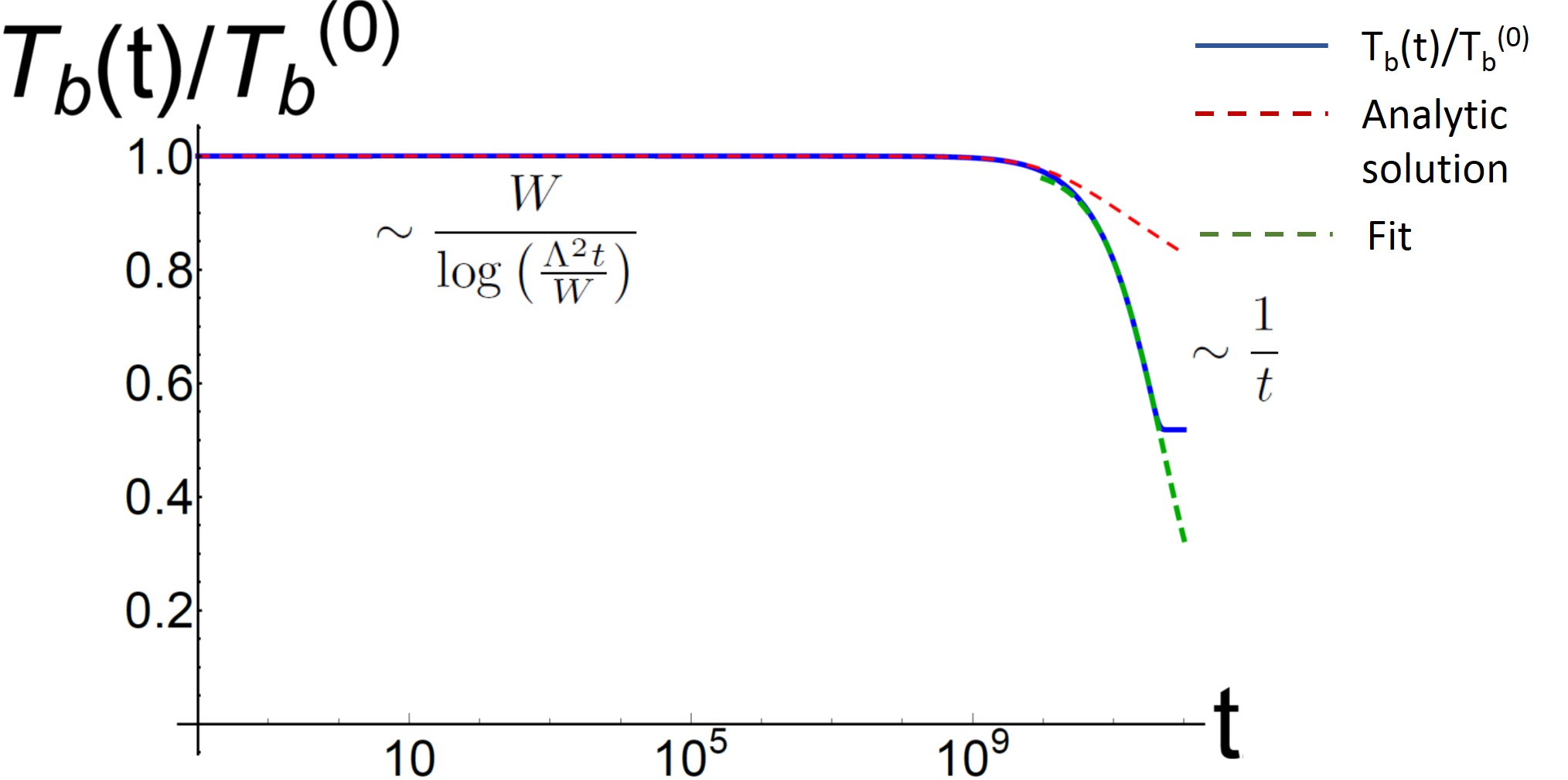}
\caption{\raggedright The time dependence of the bath $T_b(t)$ displays a $\log(t)$ behaviour over an exponentially long time scale, followed by power law cooling until equilibration. The red dashed line is the analytic approximation, Eq.~\eqref{bathtemp} to the solution of Eq.~\eqref{fractonboson}, with the numerical solution shown in blue. The green dashed line is a fit to the $1/t$ behaviour. The same parameters are used as those in Fig.~\ref{tempplot}.}
\label{tempbath}
\end{figure}

In the case where the fractons are initially prepared at a vanishing energy density,
we are in the regime where $T_f^{(0)} \ll T_b^{(0)}$. 
Here, we find that (see App.~\ref{DiffEq} for details)
\begin{align}
\label{fractontemp}
T_f(t) = -\frac{W/2}{\log\left(\frac{6 \Lambda^2}{W} t + b \right) - \frac{2W}{T_b^{(0)}}},
\end{align}
where $b = \exp\left( \frac{2W}{T_b^{(0)}} - \frac{W}{2 T_f^{(0)}} \right)$, and 
\beq
\label{bathtemp}
T_b(t) = \frac{W}{\log\left(\frac{3 \Lambda^2}{W}t + e^{W/T_b^{(0)}}\right)},
\eeq
such that the fractons display logarithmically slow heating and the bath correspondingly cools logarithmically slowly.
This behaviour persists until the fractons are close to equilibration,
i.e., over an exponentially long time scale
\beq
0 \leq t \lesssim t^* = \frac{W}{6 \Lambda^2} \exp\left(\frac{W}{T_b^{(0)}}\right).
\eeq
Until this time scale, the dominant processes are those of channels 1 and 2, where two bosons combine to pump energy into the fracton sector. Beyond this, however, channel 3 is activated since the fractons are close to equilibration ($T_f(t^*)\sim T_b^{(0)}/2$) while the bath's temperature has only changed slightly ($T_b(t^*)\sim T_b^{(0)}$), such that $n_f(t^*) \sim n_b(t^*)$. Hence, we can no longer ignore processes where a boson and a fracton convert into three fractons and the behaviour of the bath is modified at this time scale, 
\beq
T_b(t) \sim \frac{W^2}{\Lambda^2 t} e^{W/T_b^{(0)}}, \quad t>t^* .
\eeq
Thus, the bath first cools logarithmically over an exponentially long time scale, and then rapidly equilibrates with power-law behaviour, since the logarithmic heating of the charged fracton sector establishes a finite density of fractons at $t \sim t^*$. This behaviour can be explicitly seen by numerically solving Eq.~\eqref{fractonboson} (see Fig.~\ref{tempplot}) and matching the analytic solutions with the numerical curves. As can be seen in Fig.~\ref{tempfracton}, the fractons display logarithmic heating essentially until equilibration while the bath displays logarithmic cooling followed by rapid, power-law cooling, as shown in Fig.~\ref{tempbath}. 

Thus, placing a type I fracton model prepared in its ground space\footnote{We note that while we are primarily interested in the limit $T_f^{(0)} \ll T_b^{(0)}$, due to the form of Eq.~\eqref{fractonboson}, our discussion holds more generally as long as $T_f^{(0)}<T_b^{(0)}/2$.} in contact with a finite temperature heat bath leads to equilibration that exhibits glassy behaviour (i.e. a $\log t$ approach to equilibrium) over an exponentially wide window of intermediate time scales,
\beq
0 \leq t \lesssim \frac{c W}{\Lambda^2} \exp\left(\frac{W}{T_b^{(0)}}\right),
\eeq
for some positive constant $c$.

At long times, the system eventually equilibrates, $T_b = T_f$ and close to equilibration, we recover the standard exponential relaxation expected from Newton's law of cooling,
\beq
\delta T (t) \sim \exp \left(-\frac{\Lambda^2}{W} e^{-\frac{W}{T_b^{(0)}} }t\right),
\eeq
where $\delta T = T_b - T_c$. Thus, over exponentially long time scales, governed by $W/T_b^{(0)}$, the relaxation displays logarithmic (glassy) dynamics but close to equilibrium we recover an exponential relaxation with a relaxation rate that is exponentially small in $W/T_b^{(0)}$. 

\subsubsection{Fracton dynamics in an open quantum system}

In principle, we can also consider an open quantum system such that the temperature of the composites (and so also the dim-2 bosons) is pinned to that of an external heat bath. In this situation, we are interested in the dynamics of the charged fracton sector which, following the discussion in the previous system, we expect will again demonstrate logarithmically slow heating over exponentially long time scales. 

\begin{figure}[t]
\centering
\includegraphics[width=8cm]{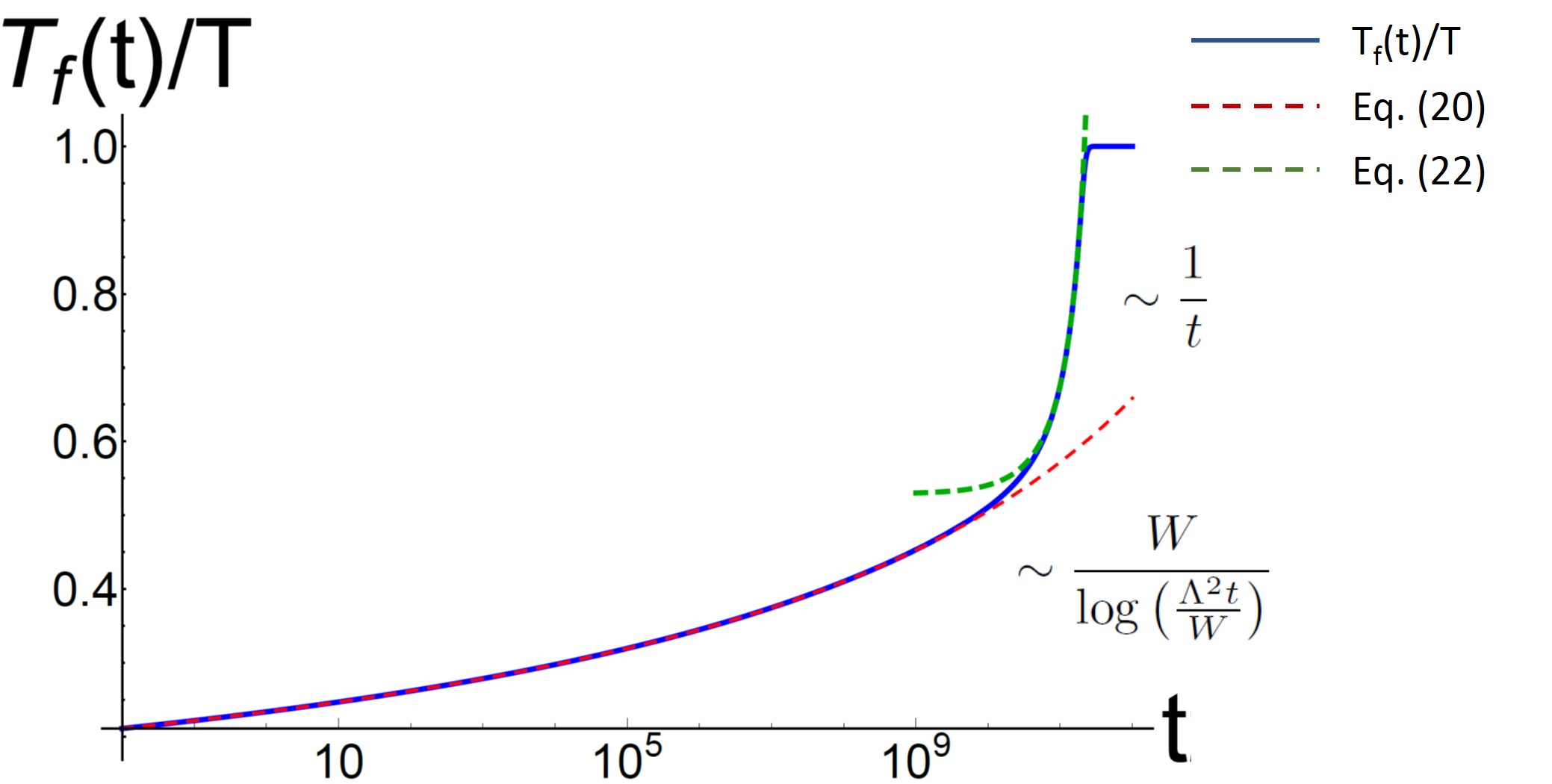}
\caption{\raggedright In an open quantum system the fractons $T_f(t)$ display a $\log(t)$ behaviour over an exponentially long time scale, followed by power law heating until equilibration. The red dashed line is the analytic approximation, Eq.~\eqref{fractonlog} to the solution of Eq.~\eqref{openfracton}, with the numerical solution shown in blue. The green dashed line depicts the $1/t$ behaviour, Eq.~\eqref{fractonpower}. The parameters are $W=1$, $\Lambda = 0.1$, $T = 0.05$, and $T_f^{(0)} = 10^{-3}$.}
\label{fractonheating}
\end{figure}

We consider initializing the charged fracton sector at a temperature $T_f^{(0)}$, with the neutral composites and dim-2 bosons coupled to an external bath at a temperature $T \gg T_f^{(0)}$. Since the fracton sector only exchanges energy with the composites and bosons, most of the discussion from the previous section holds i.e., the dominant processes by which the fracton sector exchanges energy are unchanged. Hence, the rate equation for the fractons is 
\beq
\label{openfracton}
\frac{d T_f}{d t} = \frac{4\Lambda^2 T_f^2}{W^2} \left(3 \frac{n_b^2}{n_f} + n_b - n_b n_f - n_f^2 - 2 n_f^3\right),
\eeq  
where $n_b = e^{-W/T}$ is set by the external heat bath
and $n_f = e^{-W/2 T_f}$. 
In the regime of interest, $T_f^{(0)} \ll T$, 
the dynamics are initially governed by channels 1 and 2, 
such that the fractons display logarithmically slow heating (see App.~\ref{DiffEq} for details)
\beq
\label{fractonlog}
T_f(t) = -\frac{W/2}{\log\left(\frac{6\Lambda^2}{W}t + b\right) - \frac{2W}{T}},
\eeq
where $b = \exp \left( \frac{2W}{T} - \frac{W}{2 T_f^{(0)}} \right)$. 
This behaviour persists over an exponentially long time scale, 
\beq
0 \leq t \lesssim t^* = \frac{W}{6\Lambda^2}\exp\left(\frac{W}{T}\right),
\eeq
controlled by the temperature of the heat bath, $T \ll W$. Around $t = t^*$, however, channel 3 is activated since a finite density of fractons has been established, and the behaviour of the fractons is modified, 
\beq
\label{fractonpower}
T_f(t) \approx -\frac{W/2}{\frac{2\Lambda^2}{W}e^{-W/T}\,t + \log\left(3 e^{-W/T}\right)},\quad t > t^* .
\eeq
Hence, the fractons display logarithmically slow heating over an exponentially long time scale, followed by rapid power-law behaviour until they are close to equilibration (see Fig.~\ref{fractonheating}). Near equilibration, $T_f \sim T$, and the fractons follow the usual exponential behaviour expected from Newton's law
\beq
\delta T \sim \exp\left(-\frac{4\Lambda^2 t}{W}e^{-W/T} \right),
\eeq
where $\delta T = T - T_f$. Thus, even in an open quantum system, fractons display glassy behaviour over exponentially long time scales when prepared in their ground state, where the time scales are controlled by $W/T$ i.e., by the initial temperature of the heat bath. In principle, our discussion applies as long as $T_f^{(0)}< T/2$, in which case the width of the $\log(t)$ behaviour will depend on the initial temperature of the fractons, but the situation considered here $T_f^{(0)} \ll T$ is of more interest.

While we have focused on the example of the X-Cube model Eq.~\eqref{HXC} here, we expect that our results should hold in general for Type I fracton models. In particular, our results here are consistent with the glassy behaviour demonstrated for the related CBLT model~\cite{Chamon2005}, where a Type I fracton system initialized with an isolated set of fractons and coupled to an external bath was shown to display logarithmic relaxation. While that work considered fractons which were coupled directly to the bath, in contrast we prepare our system with a finite density of fractons which are coupled to neutral composites and dim-2 excitations held at a fixed temperature $T$. 

\subsection{Role of $m^{(1)}$ excitations}
\label{m-excitations}

Type I models, such as the X-Cube model Eq.~\eqref{HXC}, may host additional excitations created in pairs at the ends of Wilson lines of $\sigma_x$ operators. These excitations are free to move in one-dimension and hence referred to as $m^{(1)}$ excitations. In order to study the X-Cube model in full generality, we must thus consider the model in the presence of transverse fields, Eq.~\eqref{HXCfull}
\beq
H = - J \sum_c A_c - \sum_{v,k} B_c^{(k)} + \Lambda \sum_i \sigma_z + \lambda \sum_i \sigma_x 
\eeq
where the term $\lambda \sigma_x$ introduces dynamics for the $m^{(1)}$ excitations. Importantly, since the $m^{(1)}$ excitations are created by $\sigma_x$ operators, they do not inter-convert with the $e$ excitations (composites, dim-2 bosons, and fractons). Thus, the only coupling between the $e$'s and $m$'s will be through the exchange of energy between the two sectors.

Let us first consider the case where $J = 1$ with $\lambda,\Lambda \ll W$. Acting with a single $\sigma_x$ operator violates four vertex terms, $B^{(k)}$, such that the gap for creating a single $m^{(1)}$ is equal to $W$, which is the same as that of creating a dim-2 boson. For a system in equilibrium at some temperature $T$, the density of $m^{(1)}$'s will thus be $n_m \sim e^{-W/T}$. In addition to the relaxation channels considered in Sec.~\ref{finiteenergy}, the finite density of $m^{(1)}$'s will then provide an additional relaxation channel for the fractons, since they can now borrow energy from the one-dimensional bath of $m^{(1)}$'s with a bandwidth controlled by the hopping rate $\lambda$. 

Since the ``magnetic" bath can supply a maximum energy $\tilde{\lambda} = \min(\lambda,T)$, the relaxation rate of the fractons due to coupling to this channel will proceed at a rate $\Gamma \sim n_f e^{-W/\tilde{\lambda}}$. If $T < \lambda$, then the relaxation will proceed at a rate controlled by the temperature $T$ and $\lambda$ will be rendered irrelevant. On the other hand, if $T > \lambda$, then it is more efficient for the fractons to couple to the ``electric" bath, since boson mediated hopping (see Fig.~\ref{bosonhop}) proceeds at a faster rate $\sim n_f e^{-W/T}$. Hence, including the dim-1 excitations does not effect the relaxation of fractons, at least in equilibrium, and can safely be ignored insofar as $\lambda$ is weak enough to not destroy the fracton topological order.

As long as $J = 1$, the gap for creating an $m^{(1)}$ and an $e^{(2)}$ is the same. Thus, even if we were to consider a system in non-equilibrium, there will be an efficient equilibration between the dim-2 bosons and the dim-1 magnetic excitations, since processes where two $m^{(1)}$'s are destroyed and two $e^{(2)}$'s are simultaneously created will proceed in an on-shell manner, with the rate of equilibration controlled by the relative strength of the transverse fields $\lambda,\Lambda$. Once these sectors have rapidly equilibrated, we can again ignore the $m^{(1)}$'s since the most dominant processes through which the fractons equilibrate will be those considered in Sec.~\ref{fractonglass}.

If $J \neq 1$, however, then there will exist an imbalance between the electric and magnetic gaps and the equilibration between the $e^{(2)}$'s and $m^{(1)}$'s will no longer proceed in an on-shell manner. Since the case where $J \neq 1$ may also de-stabilise the fracton topological order, depending on the perturbation strengths, we leave the dynamics of the X-Cube model in this case as an open question, to be studied once the phase diagram of this model is better understood \cite{Vijay2017,Ma2017}.

\section{Type II Fracton Models}
\label{type2}

We now turn to type II fracton models, 
such as Haah's code~\cite{Haah2011}. 
The fundamental difference between type I and II models 
is the lack of any local string-like operator 
that allows topological excitations to move in the latter. 
As we saw in the type~I case, pairs of fractons form dimension-1 topological excitations, 
and as pairs are created at the ends of Wilson-lines, 
they can move along the Wilson lines without creating any further excitations. 
However, in type~II models, there are no mobile subdimensional excitations. 

\begin{figure}[t]
\centering
\includegraphics[width=8cm]{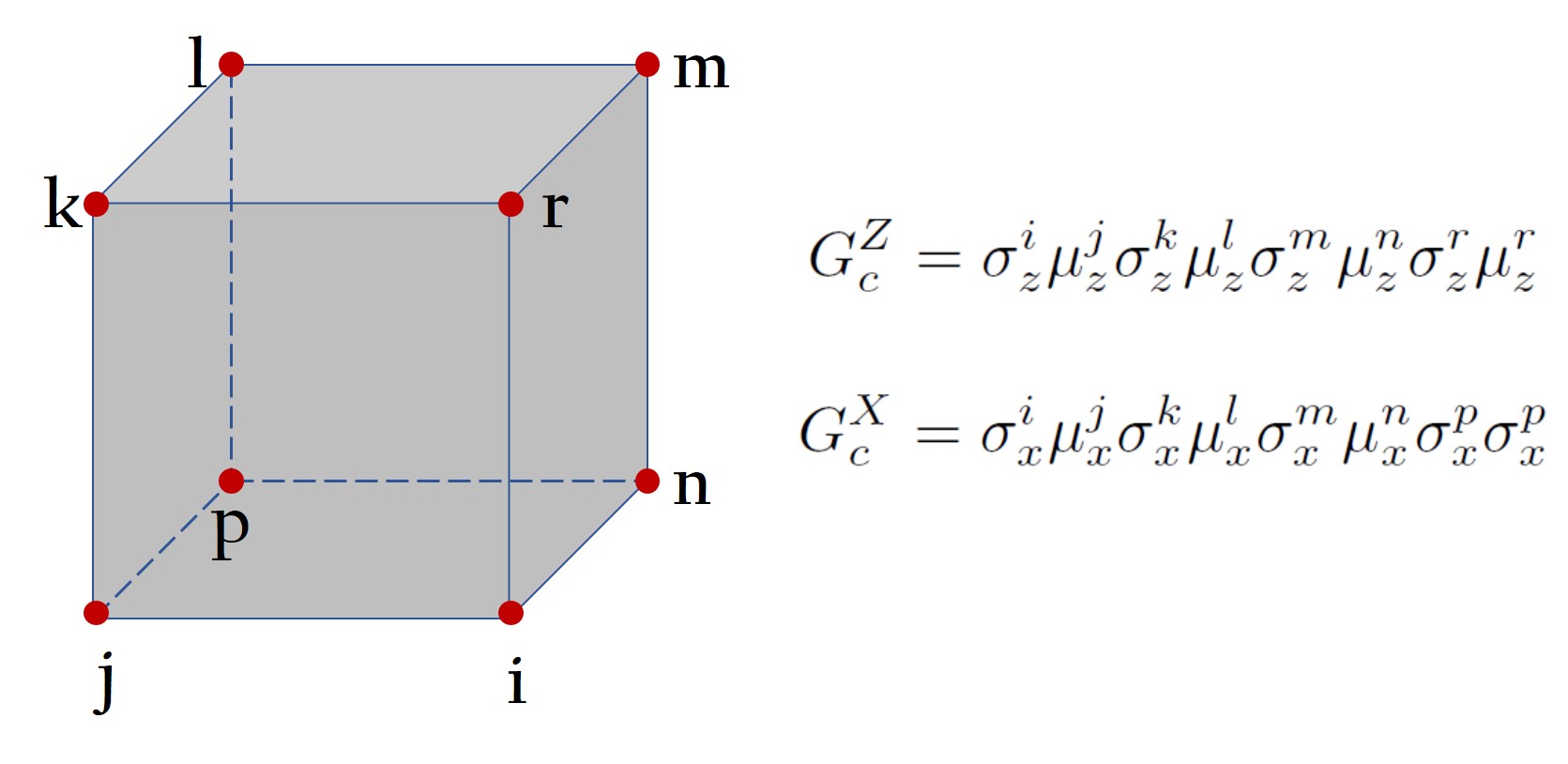}
\caption{\raggedright The Cubic Code model is defined on a cubic lattice with two spins (qubits) living on each vertex. The Hamiltonian is a sum of two eight-spin operators on each cube $c$.}
\label{cubiccode}
\end{figure}

\begin{figure}[t]
\centering
\begin{subfigure}[b]{0.22\textwidth}
\includegraphics[width=\textwidth]{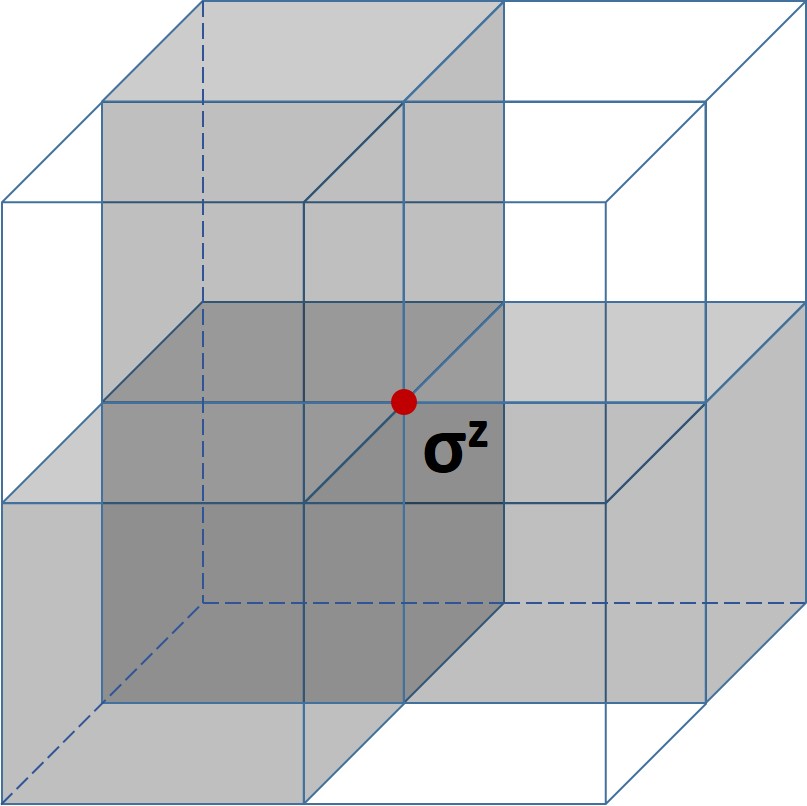}
\caption{}
\label{cubic-1}
\end{subfigure}\quad\quad
\begin{subfigure}[b]{0.22\textwidth}
\includegraphics[width=\textwidth]{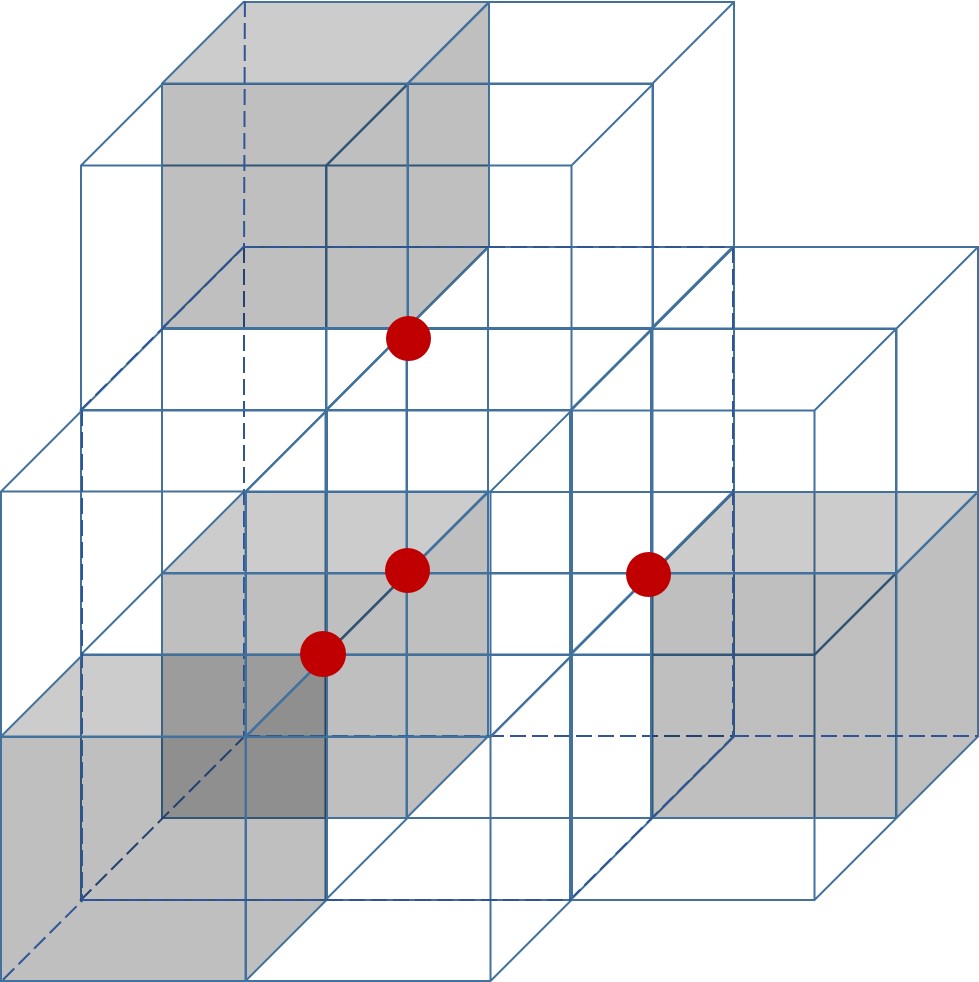}
\caption{}
\label{cubic-2}
\end{subfigure}
\caption{\raggedright Excitations of the Cubic Code Model. Two spins (qubits) live on each vertex of the cubic lattice. Acting by a $\sigma_z$ operator on the ground state, as shown in (i), creates four fracton excitations. Four fracton excitations are also created by acting by a $\mu^z$ operator or by violating the $G_c^X$ term. Acting by a fractal operator of $\sigma_z$'s, as in (ii), separates the fractons.}
\end{figure}

As a specific example of a type~II fracton phase, 
let us consider Haah's cubic code model. 
This is an exactly solvable lattice model 
defined on a three-dimensional cubic lattice 
where each site has two spins (or qubits) placed on it.
The Hamiltonian is given by the sum over all cubes of two eight-spin interaction terms
\beq
H = - J \sum_c G_c^X - \sum_c G_c^Z
\eeq
with $G_c^X$ and $G_c^Z$ defined in Fig.~\ref{cubiccode}.
In analogy with the X-Cube model, Eq.~\eqref{HXC}, 
we set $J = 1$ here, noting that once we perturb the system, 
our analysis will hold for $J$ being order unity. 
The pure Cubic Code model has a topological ground state degeneracy 
since the ground states cannot be distinguished by any local operator. 
On a three-torus of length $L$, 
the ground state degeneracy is $2^{k(L)}$ for some integer 
$2 \leq k(L) \leq 4 L$; see~\cite[Corollary~9.3]{Haah2013poly} for an explicit formula. 
There exist two kinds of excitations in this model: 
$e$ type (violation of the $G_c^X$ term) and $m$ type (violation of the $G_c^Z$ term).
This model has an exact duality between the two types 
as they are lattice inversions of each other, 
so it suffices to consider only one sector. 
We can see that these topological excitations are immobile 
since a single $\sigma_z$ operator on a link creates four cube excitations (see Fig.~\ref{cubic-1}) 
and there exists no local string-like operator 
that can create just a single pair of fractons. 
Here, single fractons are created at the ends of fractal operators, 
as shown in Fig.~\ref{cubic-2}. 
Importantly, the ``no-strings" rule~\cite{Haah2011} 
implies that \emph{any} cluster of defects with a non-trivial topological charge must be immobile.

Unlike type I fracton phases, where there is a finite and constant energy barrier for topological charges to move, in type II fracton phases there exists an extensive logarithmic energy barrier preventing topologically charged excitations from diffusing. In particular, as proved in~\cite{Bravyi2011b}, there exists an energy barrier $\sim c \log(R)$ for creating an isolated fracton with no defects within a distance $R$, for some constant $c$.

Let us consider a type II fracton system in contact with a narrow bandwidth heat bath of composites, with bandwidth $\Lambda \ll W$ and temperature $T$. Let us consider a (non-equilibrium) initial state containing a single fracton.  Following our logic in the type I case and replacing the energy scale $W$ in the charged sector by the new energy scale $W c \log R$, we conclude that to move a distance $R$ will take a time 
\beq
t = \left(R \right)^{\frac{c W}{\tilde \Lambda}} ;\quad  \tilde \Lambda = \min(T, \Lambda).
\eeq
Inverting this relation we obtain $R \sim t^{\tilde \Lambda/c W}$, which for $\tilde \Lambda \ll W$ implies strongly {\it subdiffusive} behaviour in a translation invariant three-dimensional model. This subdiffusive behaviour will persist up to the relaxation time, which (as we will shortly show) is super-exponentially long at low temperatures, scaling as $t_{relax} \sim  \exp(+ c' \frac{W^2}{T^2})$. 

Of course, a type II system left in contact with a heat bath will eventually equilibrate to have a non-zero density of fractons---fractons will be created in groups of four (borrowing the energy to do this from the heat bath), and will then be redistributed over the system. However, to achieve an equilibrium distribution with a thermal density of fractons $n_f \sim \exp(-W/2T)$, it will be necessary to move fractons over a lengthscale 
$a \sim \exp(2W/3T)$ (given that fractons can only be created in groups of four). 
Substitution into the above equation then leads us to conclude that the equilibration time 
will follow the ``super-Arrhenius" scaling 
\begin{equation}
t_{equilibrate} \sim \exp \left(+ c' \frac{W^2}{T^2} \right), \label{superarrhenius}
\end{equation}
when $T < \Lambda$, where $c'$ is an $O(1)$ numerical constant. This is consistent with a lower bound on relaxation rates derived in~\cite{BravyiHaah2013}. 
This ``super-Arrhenius" scaling is reminiscent of various ``near MBL" models 
e.g.~\cite{GN, MirlinMuller} and provides another example of the provocative connections between fracton dynamics and MBL. 

We note that classical spin systems, 
such as the Newman-Moore model~\cite{Newmanmoore}, 
have been known to display similar phenomenology, 
in that there exists a logarithmic energy barrier for transporting defects, 
leading to glassy behaviour. 
However, unlike type II fracton models, these classical spin models 
do not have a topologically ordered ground-state subspace
and hence lack the robustness against generic local perturbation 
inherent in quantum fracton models, such as Haah's code~\cite{BravHast2010,BravHast2011}. 

We now compare our results to~\cite{SivaYoshida}, 
who argued that the memory of the initial conditions in the cubic code relaxes 
for long enough times at finite temperature,
with a relaxation time that scaled similarly to Eq.~\eqref{superarrhenius}.
Our broad conclusions are in agreement with~\cite{SivaYoshida}. 
However, our analysis differs in that~\cite{SivaYoshida} examined dynamics towards Gibbs states 
making use of the Lindblad formalism, which assumes decoherence, whereas we are discussing closed system quantum dynamics, with equilibration implicitly defined in terms of eigenstate thermalization \cite{ARCMP} and are making no assumptions about decoherence. 

To conclude, our analysis reveals that type II models exhibit {\it subdiffusion} of fractons 
on time scales short compared to the relaxation time, 
and additionally the relaxation time diverges 
at low temperatures as a {\it super-exponential} function of inverse temperature. 
This is reminiscent of MBL, 
in that the closed system quantum dynamics of a zero temperature fracton sector 
coupled to a finite temperature neutral sector could preserve a memory of the initial conditions 
in local observables over extremely long times 
that, while finite, are {\it super-exponentially} long at low temperature.

\section{Discussion and Conclusions}
We have shown that fracton models naturally demonstrate glassy dynamics. Type I fracton models at finite energy density exhibit a mobility that is exponentially small in the ``temperature," corresponding to a type of ``asymptotic many body localization." Meanwhile, the equilibration of type I fracton models at low temperatures involves a {\it logarithmic} approach to equilibrium over an exponentially wide range of time scales---another signature of glassy dynamics. In Type II models, individual fractons exhibit {\it subdiffusive} dynamics up to a relaxation time that is super-exponentially long at low temperature, reminiscent of various near-MBL systems, but involving a {\it translation invariant} three-dimensional Hamiltonian.  

This work has provocative implications for numerous fields. Firstly, it opens up a new direction for the investigation of three-dimensional topological order, suggesting that certain three-dimensional topologically ordered phases (the fracton models) naturally exhibit a rich glassy dynamics more conventionally associated with disordered systems. Can ideas from localization and disordered systems be fruitfully employed to understand three-dimensional topological order? Are there more surprises in store regarding the dynamical behaviour of such models? Given the novelty of these systems, it seems likely that there is more to be discovered. 

Additionally, this work may have implications for quantum foundations, in that it identifies a class of quantum systems as being unusually robust to thermalization (in which coupling to a heat bath ``observes" the system and ``collapses" the wavefunction). This robustness is particularly strong for type II models, for which the time scale for thermalization diverges super-exponentially fast at low temperatures. These models could also have important technological implications, insofar as they display a long lived memory of the initial conditions. 

There may also be a connection to the ``quantum disentangled liquids" (QDL) introduced in~\cite{Grover2014}. The hallmark of the QDL is that it contains two species of particles, and the many body eigenstates have volume law entanglement entropy, but after performing a projective measurement on the more mobile species of particles, the resulting wavefunction has only area law entanglement entropy. In the fracton models at non-zero temperature, there are indeed two species of particles (fractons and neutral composites), and the neutral composites are in a thermal state, so the eigenstates will indeed have volume law entanglement entropy. However, the fractons themselves can {\it only} hop by coupling to the neutral sector, and after performing a projective measurement on the neutral sector, the ``fracton only" portion of the wavefunction may well have sub-volume law entanglement entropy. If so, then finite temperature fracton models would provide a three-dimensional realization of a quantum disentangled liquid. Unfortunately, a direct numerical test of this scenario seems difficult, since these models are intrinsically three-dimensional, and numerical tools capable of dealing with three-dimensional glassy many body systems are severely limited.

Thus far we have worked with models with only a $Z_2$ charge. However, fracton models should admit of generalizations 
with $U(1)$ conserved charge~\cite{Pretko1, Pretko2} 
allowing us to define a charge {\it conductivity}. 
It should then follow, through reasoning analogous to our previous discussion of type II fracton models, 
that at zero temperature (but with the neutral composites prepared at finite density), 
that these models should realize a phase that is a {\it thermal} conductor but a {\it charge} insulator. 

Finally, there are the implications for the study of MBL and glass physics. We have introduced a new class of translation invariant models which naturally exhibit glassy dynamics in three dimensions. This new class of models may well provide a new line of attack on important unsolved problems such as MBL in translation invariant systems~\cite{MaksimovKagan, Grover2014, Schiulazbubbles, yaoglass, Papicstoudenmire, deroeck1, deroeck2, Garrahanglass} and in higher dimensions~\cite{lstarbits, avalanches}. Finally, while the fracton models are defined on lattices, analogous phenomenology can also be obtained in the continuum by making use of higher rank gauge theories. Given the interest in MBL in the continuum~\cite{2dcontinuum, gornyicontinuum}, these too may be worthy of investigation. We leave exploration of these various issues to future work. 

\begin{acknowledgments}
We acknowledge useful conversations with Liang Fu, Mike Hermele, Han Ma, and Michael Pretko. 
RN is supported by the Foundation Questions Institute under grant number FQXi-RFP-1617.
\end{acknowledgments}

\appendix

\section{Details on Type I Relaxation Dynamics}
\label{DiffEq}

Here, we discuss in detail the solutions of the relevant differential equations encountered in the main text. All of the relevant equations can be brought to the form
\beq
\label{maineq}
\frac{dy}{dt} = a y^2 e^{\pm \frac{1}{y}}.
\eeq
Since this is a separable equation, this is equivalent to
\beq
\int_{y(0)}^{y(t)} \frac{dz}{z^2} e^{\mp \frac{1}{z}} = a t.
\eeq 
Changing variables to $x = \pm \frac{1}{z}$, we get
\beq
\mp \int_{\pm 1/y(0)}^{\pm 1/y(t)} e^{-x} = a t. 
\eeq
Thus, the solution is 
\beq
y(t) = \frac{\mp 1}{\log\left(e^{\mp 1/y(0)} \pm a t \right)}
\eeq

\subsection{Equilibration within the bath}
\label{bathdetails}

\begin{figure}[b]
\centering
\includegraphics[width=8cm]{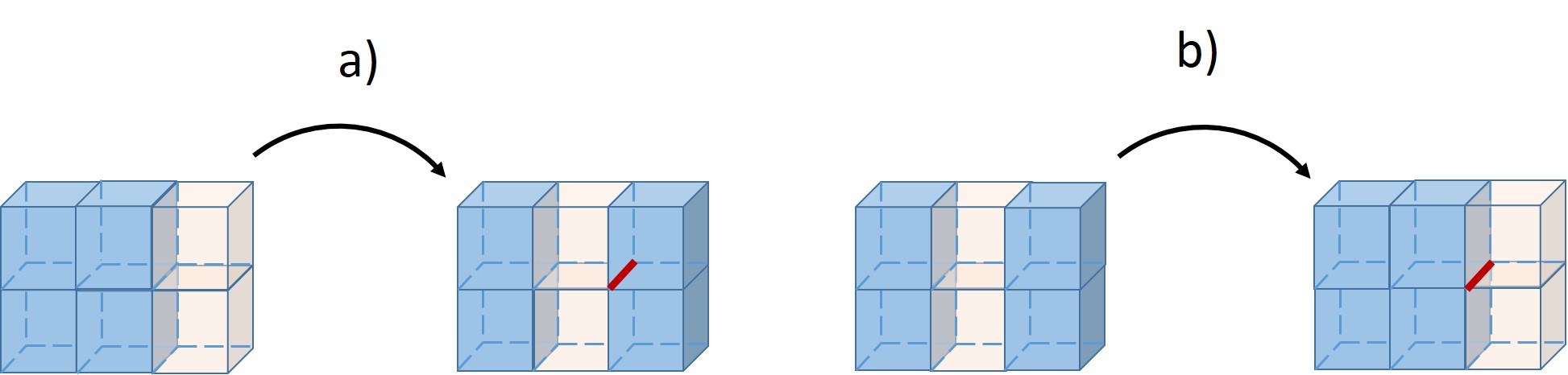}
\caption{\raggedright a) Acting by a single $\sigma_z$ operator decomposes a neutral composite (bosons) into two dim-2 excitations (bosons). b) The reverse process, where two $e^{(2)}$'s combine into a single composite.}
\label{bosoncomposite}
\end{figure}

First, we consider the dynamics within the bath, ignoring the fractons. 
As depicted in Fig.~\ref{bosoncomposite}, a composite can decompose into two 
bosons while remaining on-shell. Being a first order process 
(in $\Lambda \ll W$), this occurs at a rate $\sim \Lambda n_c$.
Similarly, the reverse process where two bosons combine to form a composite occurs 
at a rate $\sim \Lambda n_b^2$. To leading order, these are the only two 
processes which contribute to the equilibration between these two sectors, 
as all other processes require additional composites/bosons and are suppressed 
by additional density factors. Considering only the leading order processes, 
a detailed balance is established between these two sectors,
\begin{align}
\frac{d E_c}{d t} &= -2 W \Lambda (n_c - n_b^2) &= -2 W \Lambda \left(e^{-\frac{2W}{T_c}} - e^{-\frac{2W}{T_b}}\right)\nonumber \\
\frac{d E_b}{d t} &= -2 W \Lambda (n_b^2 - n_c) &= -2 W \Lambda \left(e^{-\frac{2W}{T_b}} - e^{-\frac{2W}{T_c}}\right)
\end{align}
Putting in the heat capacities Eq.~\eqref{heatcap}, we get 
\begin{align}
\label{compboseeqn}
\frac{d T_c}{d t} &= - \frac{\Lambda}{2W} T_c^2 \left(1 - e^{2W \left(\frac{1}{T_c} - \frac{1}{T_b} \right)} \right), \nonumber \\
\frac{d T_b}{d t} &= - \frac{2\Lambda}{W} T_b^2 \left(1 - e^{2W \left(\frac{1}{T_b} - \frac{1}{T_c} \right)} \right) e^{-\frac{W}{T_b}}.
\end{align}
In the regime where $T_b^{(0)} \ll T_c^{(0)}$, the bath is initially comprised primarily 
of neutral composite excitations, and in this limit the above equations are approximately 
\begin{align}
\frac{d T_c}{dt} &\approx - \frac{\Lambda}{2W} T_c^2 \, ,\nonumber \\
\frac{d T_b}{dt} &\approx \frac{2 \Lambda}{W} T_b^2 e^{\frac{W}{T_b}} e^{-\frac{2W}{T_c}}\, .
\end{align}
For $T_c(t)$, with initial condition $T_c(0) = T_c^{(0)}$, we find
\beq
\label{bathcooling}
T_c(t) = \frac{2 W}{\Lambda t + \frac{2 W }{T_c^{(0)}}}.
\eeq
Since we are working in the limit $T_c^{(0)} \ll W$, $T_c(t) \sim T_c^{(0)}$ for a time $t \sim \frac{W}{\Lambda T_c^{(0)}}$. We can hence treat $T_c(t)$ as a constant over this time scale such that $T_b(t)$ satisfies
\beq
\frac{d T_b}{dt} \approx \frac{2 \Lambda}{W} T_b^2 e^{\frac{W}{T_b}} e^{-\frac{2W}{T_c^{(0)}}}.
\eeq
This is equivalent to Eq.~\eqref{maineq} with $y = T_b/W$ and $a = 2\Lambda e^{-\frac{2W}{T_c^{(0)}}}$, and hence has the solution
\beq
\label{bosonfit}
T_b(t) = \frac{-W}{\log\left(2 \Lambda t +b \right) - \frac{2W}{T_c^{(0)}}},
\eeq
where $b = e^{\frac{2W}{T_c^{(0)}} - \frac{W}{T_b^{(0)}}}$.
This behaviour is expected to hold until the bosons are close to equilibration $T_b \sim T_c^{(0)}/2$, i.e., over the 
window
\beq
0 \leq t \lesssim \frac{k}{2\Lambda}
\eeq
where $k$ is some positive constant.
\begin{figure}[t]
\centering
\includegraphics[width=8cm]{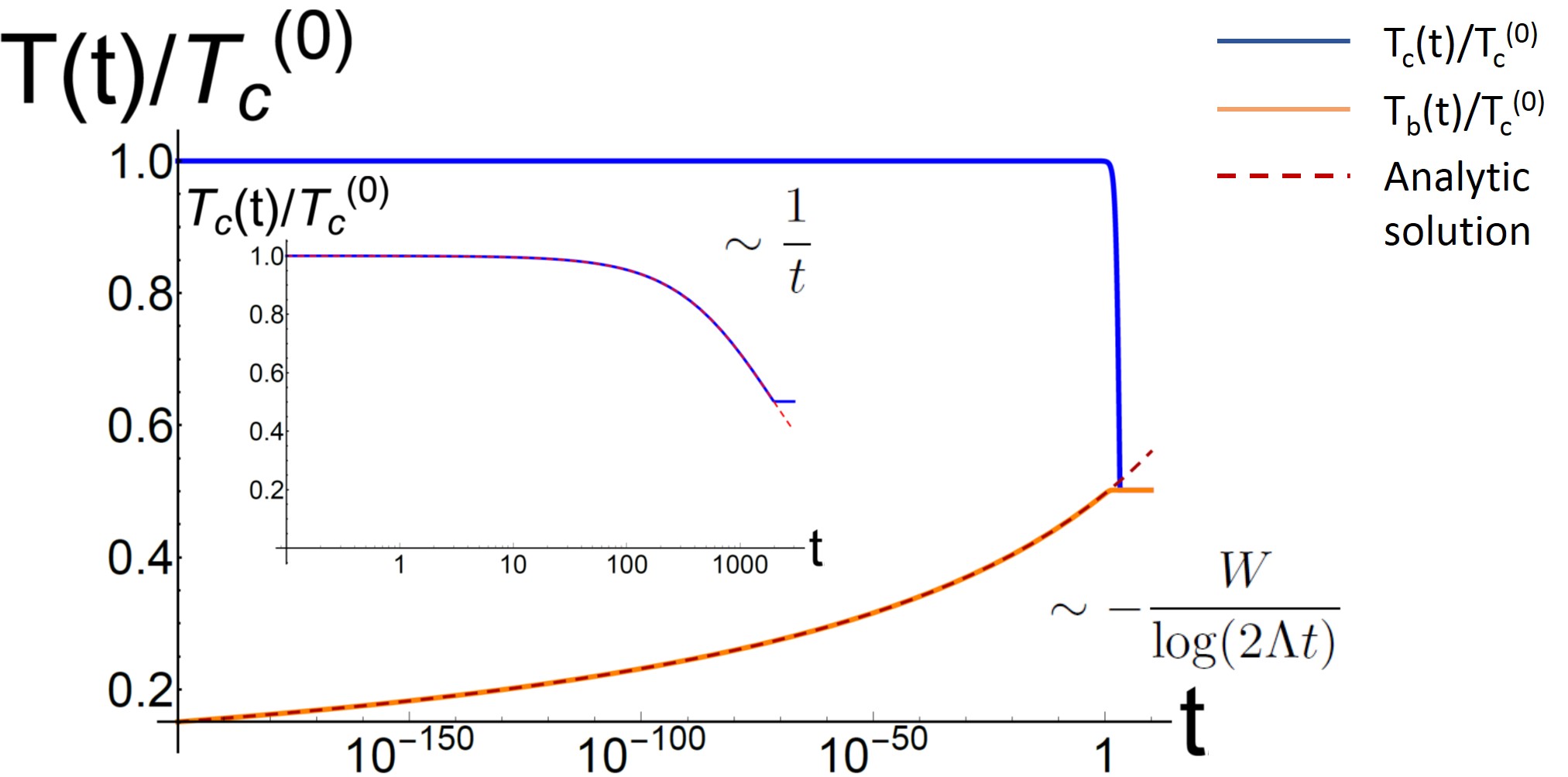}
\caption{\raggedright
The composite sector (blue) cools with power-law behaviour while the bosonic sector (orange) displays a $\log(t)$ behaviour over a short time scale, set by $\sim 1/\Lambda$. 
The red dashed line, Eq.~\eqref{bosonfit}, is the 
approximate analytic solution to Eq.~\eqref{compboseeqn} 
in the regime of interest $T_b^{(0)} \ll T_c^{(0)}$, 
shown here against the numerical solution for $T_b(t)$. 
The in-set compares the analytic $1/t$ behaviour Eq.~\eqref{bathcooling} (red dashed line) 
with the numerical solution for the composite sector (blue). 
The parameters here are $W=1, \Lambda = 10^{-1}$, $T_c^{(0)} = 10^{-2}$, and $T_b^{(0)} = 0.5 \times 10^{-3}$.}
\label{fastcooling}
\end{figure}

\begin{figure}[t]
\centering
\includegraphics[width=8cm]{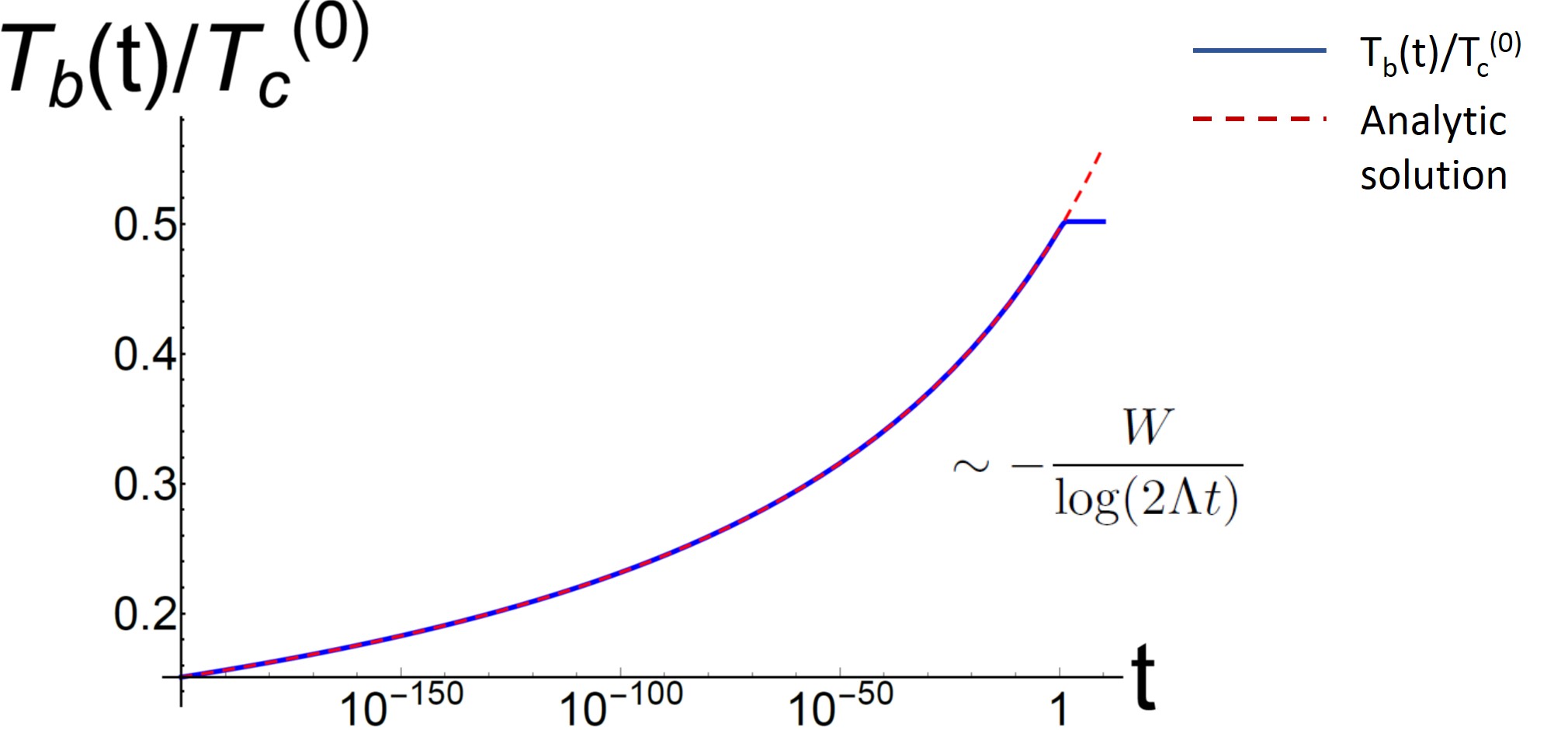}
\caption{\raggedright The bosonic sector displays a $\log(t)$ behaviour over a short time scale, set by $\sim 1/\Lambda$. The red dashed line is the approximate analytic solution, Eq.~\eqref{bosonfit}, for Eq.~\eqref{compboseeqn} shown against the numerical solution for $T_b(t)$, in blue. The parameters here are $W=1, \Lambda = 10^{-1}, T_c^{(0)} = 10^{-2}$, and $T_b^{(0)} = 0.5 \times 10^{-3}$.}
\label{bosonlogfit}
\end{figure}

The bosons thus display logarithmic heating over a time scale $0 \leq t \lesssim \frac{k}{\Lambda}$, for some positive constant $k$. They reach the equilibrium temperature faster than the neutral composites, which cool down with power-law behaviour over a time scale $0\leq t \lesssim \frac{k' W}{\Lambda T_c^{(0)}}$ governed by their initial temperature, $T_c^{(0)}$. The equilibration process is depicted in Fig.~\ref{fastcooling}, 
where the approximate analytic solutions developed above 
are shown to be in good agreement with the numerical solutions for Eq.~\eqref{compboseeqn}. 
In particular, as illustrated in Fig.~\ref{bosonlogfit}, 
the approximate analytic solution Eq.~\eqref{bosonfit} fits the numerical solution for Eq.~\eqref{compboseeqn} quite well,
confirming the logarithmic heating of the bosonic sector. 

Close to equilibration, we set $T_c = T + \frac{\delta T}{2}$ and $T_b = T - \frac{\delta T}{2}$, such that Eq.~\eqref{compboseeqn} leads to 
\beq
\frac{d (\delta T)}{d t} \approx - \Lambda T,
\eeq
in the limit $\delta T \ll T$ and $T = T_c^{(0)}/2 \ll W$. Thus, close to equilibration the system recovers an exponential relaxation rate since $\delta T(t) = T_c(t) - T_b(t)$ behaves according to Newton's law of cooling: $\delta T(t) \sim e^{-\Lambda t}$.
 
The above discussion demonstrates that while we could, in principle, initialize our system 
without any bosonic excitations, the presence of a finite density bath of 
neutral composites will quickly establish a finite density of bosons 
in equilibrium with the composites (see Fig.~\ref{fastcooling}). Following 
this discussion, it is easy to show that starting in the opposite limit i.e., 
with $T_c^{(0)}\ll T_b^{(0)}$ will result in the same equilibrium configuration, 
except with the dynamics reversed: the bosons will cool with a power law while the composites will 
display a $\log(t)$ behaviour. When considering the dynamics of the fracton sector, 
it is therefore reasonable to assume that the composite and bosonic sectors have 
already equilibrated. Thus, we make this assumption in the main text and consider 
a heat bath with both the composites and bosons at an initial temperature $T_b^{(0)}$.

\subsection{Equilibration between bath and fractons}

We now consider equilibration between the charged fractons and the bath (comprised of $e^{(2)}$'s and composites). The equilibration process is governed by Eq.~\eqref{fractonboson},
\begin{align}
\label{fractonboson2}
\frac{d T_b}{d t} &=  - \frac{\Lambda^2 T_b^2}{W^2} \left(3 n_b + n_f - n_f^2 - \frac{n_f^3}{n_b} - 2 \frac{n_f^4}{n_b}\right), \nonumber \\
\frac{d T_f}{d t} &= \frac{4\Lambda^2 T_f^2}{W^2} \left(3 \frac{n_b^2}{n_f} + n_b - n_b n_f - n_f^2 - 2 n_f^3\right).
\end{align}
In the regime of interest, $T_f^{(0)}\ll T_b^{(0)}$, such that the above equations reduce to 
\begin{align}
\frac{d T_b}{d t} &\approx -3 \frac{\Lambda^2 T_b^2}{W^2} e^{-\frac{W}{T_b}}, \nonumber \\
\frac{d T_f}{d t} &\approx 12 \frac{\Lambda^2 T_f^2}{W^2} e^{\frac{W}{2 T_f}} e^{-\frac{2W}{T_b}}.
\end{align}
The first equation is equivalent to Eq.~\eqref{maineq} with $y = T_b/W$ and $a = -3 \Lambda^2/W$ and hence has the solution 
\beq
T_b(t) = \frac{W}{\log\left(\frac{3 \Lambda^2}{W}t + e^{W/T_b^{(0)}}\right)},
\eeq
establishing the logarithmic cooling displayed by the bath. To understand the heating of the fractons, we first note that the bath temperature stays roughly constant, $T_b(t) \sim T_b^{(0)}$, for an exponentially long period of time $t \sim \frac{W}{3\Lambda^2}e^{W/T_b^{(0)}}$. We can hence treat $T_b$ as a constant, so that the behaviour of the fractons is governed by 
\beq
\frac{d T_f}{d t} \approx 12 \frac{\Lambda^2 T_f^2}{W^2} e^{\frac{W}{2 T_f}} e^{-\frac{2W}{T_b^{(0)}}},
\eeq
which is again of the form Eq.~\eqref{maineq} and has the solution
\beq
T_f(t) = -\frac{W/2}{\log\left(\frac{6 \Lambda^2}{W}t + b \right) - \frac{2W}{T_b^{(0)}}},
\eeq
where $b = e^{\frac{2W}{T_b^{(0)}} - \frac{W}{2 T_f^{(0)}}}$. Since the bath cools logarithmically over an exponential time scale, this behaviour of the fractons holds until $T_f(t) \sim T_b^{(0)}/2$ i.e., over an exponentially long time scale
\beq
0 \leq t \lesssim \frac{W}{6 \Lambda^2} e^{W/T_b^{(0)}}.
\eeq
Thus far, we have only considered processes where two bosons combine to pump energy into the fracton sector, and have neglected processes where a boson and a fracton convert into three fractons. This is no longer accurate once the fracton temperature $T_f \sim T_b^{(0)}/2$. At this time scale $t \sim \frac{W}{6 \Lambda^2} e^{W/T_b^{(0)}}$, the fracton density $n_f \sim n_b$, since the temperature of the bath has yet to significantly deviate from its initial temperature $T_b^{(0)}$ while the fracton temperature has almost reached its equilibrium value, $T_b^{(0)}/2$. Once the fractons are close to equilibration, the behaviour of the bath is hence governed by
\beq
\frac{dT_b}{dt} \approx - \frac{\Lambda^2 T_b^2}{W^2} (3 n_b + n_f),
\eeq
where we are dropping terms of $O(n_f^2)$ and higher. The second term on the r.h.s $\sim n_f$ corresponds to channel 3, which is now activated. Treating $n_f$ as a constant (since the fracton sector is close to equilibration), we find that the bath's behaviour is now modified,
\beq
T_b(t) \sim \frac{W^2}{\Lambda^2 t} e^{W/T_b^{(0)}}.
\eeq

To study the behaviour close to equilibration, we set $T_b = T + \frac{\delta T}{2}$ and $T_f = T - \frac{\delta T}{2}$, which leads to 
\beq
\frac{d}{d t}\delta T = -\frac{\Lambda^2}{W}e^{-W/2T} \delta T,
\eeq
in the limit $T = T_b^{(0)}/2 \ll W$ and where we are interested in the infinite time i.e, $\delta T \ll T$ behaviour. Here, we recover the standard exponential relaxation expected from Newton's law of cooling
\beq
\delta T (t) \sim \exp \left(-\frac{\Lambda^2 t}{W} e^{-W/T_b^{(0)}}\right).
\eeq

\subsection{Equilibration between fractons and external heat bath}

Finally, we consider the situation where the fractons are prepared at an initial temperature $T_f^{(0)} \ll T$, where $T$ is the temperature of an external heat bath, to which the temperatures of the composites and dim-2 bosons are pinned. In this scenario, the behaviour of the fractons is governed by Eq.~\eqref{openfracton}
\beq
\label{openfracton2}
\frac{d T_f}{d t} = \frac{4\Lambda^2 T_f^2}{W^2} \left(3 \frac{n_b^2}{n_f} + n_b - n_b n_f - n_f^2 - 2 n_f^3\right),
\eeq  
where $n_b = e^{-W/T}$ and $n_f = e^{-W/2 T_f}$. As we are interested in the dynamics of fractons prepared in the ground state, in the regime of interest $T_f^{(0)} \ll T$ we find that 
\beq
\frac{d T_f}{d t} \approx \frac{12\Lambda^2 T_f^2}{W^2} e^{-2W/T} e^{W/2T_f}.
\eeq
Thus, the fractons display logarithmic heating 
\beq
T_f(t) = -\frac{W/2}{\log\left(\frac{6\Lambda^2}{W}t + b\right) - \frac{2W}{T}},
\eeq
where $b = e^{\frac{2W}{T} - \frac{W}{2 T_f^{(0)}}}$. Note that this is the same behaviour encountered in the previous section. However, once $T_f(t) \sim T/2$, we can no longer ignore processes where a boson and a fracton convert into three fractons (channel 3) since $n_b^2 n_f \sim n_b$ at this point. This is in contrast to the previous section, where the fractons had almost equilibrated by the time channel 3 was activated, leading to a logarithmic heating of fractons essentially over the entire equilibration time scale.

Hence, when the composites and dim-2 excitations are coupled to an external heat bath, the logarithmic behaviour of the fractons persists until $T_f(t) \sim T/2$, i.e., over an exponentially long time scale
\beq
0 \leq t \lesssim \frac{W}{6\Lambda^2}e^{W/T}
\eeq
controlled by the temperature of the heat bath, $T \ll W$. Beyond this time scale, however, channel 3 is active, and the dynamics of the fractons are governed by 
\beq
\frac{d T_f}{d t} \approx \frac{4\Lambda^2 T_f^2}{W^2} \left(3 \frac{n_b^2}{n_f} + n_b \right),
\eeq 
with the solution 
\beq
T_f(t) \approx -\frac{W/2}{\frac{2\Lambda^2}{W}e^{-W/T}\,t + \log\left(3 e^{-W/T}\right)}.
\eeq
Thus, the fractons first heat up logarithmically slowly, but once a finite density of fractons is established, they display power law heating until they are close to equilibration. Near equilibration, $T_f = T - \delta T$ with $\delta T \ll T$, and the fractons then heat according to 
\beq
\frac{d}{dt}\delta T \approx -\frac{4 \Lambda^2}{W} e^{-W/T} \delta T,
\eeq
such that the fractons follow the usual Newton's law close to equilibration,
\beq
\delta T \sim \exp\left(-\frac{4 \Lambda^2 t}{W}e^{-W/T} \right).
\eeq


\newpage 

\bibliography{library}


\end{document}